\documentclass[aps,showpacs,floatfix]{revtex4}
\usepackage{psfig}
\usepackage{color}
\def\mark{\widetilde}
\def\nn{\nonumber \\}

\def\ep{\epsilon}
\newcommand{\ba}{\begin{eqnarray}}
\newcommand{\ea}{\end{eqnarray}}
\newcommand{\be}{\begin{equation}}
\newcommand{\ee}{\end{equation}}

\newcommand{\order}[1]{{\cal O}\left(#1\right)}

\begin{document}

\title{
%
%
\[ \vspace{-2cm} \]
\noindent\hfill\hbox{\rm  } \vskip 1pt \noindent\hfill\hbox{\rm
\normalsize Alberta Thy 01-05} \vskip 10pt
\noindent\hfill\hbox{\rm  } \vskip -25pt \noindent\hfill\hbox{\rm
\normalsize UVIC-THY-05-01} \vskip 10pt
\noindent\hfill\hbox{\rm  } \vskip -25pt \noindent\hfill\hbox{\rm
\normalsize  IFT-12/2005 } \vskip 10pt
 The electromagnetic dipole operator effect on $\bar{B} \rightarrow X_s
  \gamma$ at ${\cal O}(\alpha_s^2)$}
\author{ I.~Blokland$^{1,2}$,
        A.~Czarnecki$^{1}$,
        M.~Misiak$^{3}$,
        M.~\'Slusarczyk$^{1,4}$,
        and F.~Tkachov$^{5}$ }
\affiliation{
1) Department of Physics, University of Alberta, Edmonton, AB T6G 2J1, Canada \\
2) Department of Physics and Astronomy, University of Victoria, Victoria, BC V8P 5C2, Canada \\
3) Institute of Theoretical Physics, Warsaw University, 00-681 Warszawa, Poland \\
4) Institute of Physics, Jagiellonian University, 30-059 Krak\'ow, Poland \\
5) Institute for Nuclear Research, Russian Academy of Sciences, Moscow, 117312, Russian Federation\\
}
\begin{abstract}
   The flavor-changing electromagnetic dipole operator $O_7$ gives the
    dominant contribution to the $\bar{B} \rightarrow X_s \gamma$ decay rate.
    We calculate two-loop QCD corrections to its matrix element together with
    the corresponding bremsstrahlung contributions. The optical theorem is
    applied, and the relevant imaginary parts of three-loop diagrams are
    computed following the lines of our recent $t \to X_b W$ calculation. The
    complete result allows us to test the validity of the naive non-abelianization (NNA)
    approximation
    that has been previously applied to estimate the NNLO QCD correction to
  $\Gamma(\bar{B} \rightarrow X_s \gamma)/\Gamma(\bar{B} \rightarrow X_u e \bar\nu)$.
  When both decay widths are normalized to $m^5_{b,R}$ in the same
  renormalization scheme $R$, the calculated ${\cal O}(\alpha_s^2)$ correction
  is sizeable ($\sim 6\%$), and the NNA estimate is about $1/3$ too large. On
  the other hand, when the ratio of the decay widths is written as $S \times
  m^2_{b,\overline{\rm MS}}(m_b)/m^2_{b,{\rm pole}}$, the calculated ${\cal
    O}(\alpha_s^2)$ correction to $S$ is at the level of $1\%$ for both the
  complete and the NNA results.

\end{abstract}

\pacs{13.40.Hq,14.65.Fy,12.38.Bx}

\maketitle

\section{Introduction} \label{sec:intro}

The radiative decay $b\to s\gamma$ occurs only through quantum loop effects,
similarly to the muon anomalous magnetic moment $(g_\mu-2)$ or
radiative hyperon decays \cite{Shifman:1976ge}. The largest contribution by
far to $(g_\mu-2)$ is due to electromagnetic interactions.  In
contrast, the flavor-changing hyperon ($s\to d$) and  ${\bar B}$ meson
  ($b\to s$) decay amplitudes involve heavy particles such as the $W$
boson and are consequently very rare.  In particular, the $b\to s \gamma
  $ transition can be predicted in the Standard Model with good accuracy,
and it offers a relatively low-background probe of possible new phenomena such
as supersymmetry (e.g., see Ref.~\cite{Gambino:2004mv}).

High-accuracy measurements of the  $\bar{B}\to X_s\gamma$ rate in
$B$-factories \cite{Alexander:2005cx} warrant sophisticated calculations of
high-order Standard Model contributions. Electroweak loop effects, without
which this decay would not occur, have now been studied to two-loop accuracy
\cite{ Czarnecki:1998tn, Strumia:1998bj, Kagan:1998ym, Baranowski:1999tq,
  Gambino:2000fz}.  For the precise determination of the rate, the QCD
effects are crucial and are fully known to the next-to-leading order (e.g.,
see Refs.~\cite{Buras:2002tp,Buras:2002er}).  Potentially important effects
resulting from the binding of the $b$ quark in  the $\bar{B}$ meson
were explored in Ref.~\cite{Falk:1993dh}.

At present, several groups are working  at  the determination of the
next-to-next-to-leading order (NNLO) QCD corrections.  A discussion of the
various required studies can be found in Ref.~\cite{Misiak:2003xb}. Since that
review was published, new ingredients have been provided: all the relevant
three-loop anomalous dimensions  \cite{Gorbahn:2004my,Gorbahn:2005sa},
three-loop matching  conditions  \cite{Misiak:2004ew},  as well as
certain counterterm contributions to the three-loop matrix elements of
four-quark operators \cite{Asatrian:2005pm} .

Among the most challenging missing quantities are the two-  and
  three-loop matrix elements of several operators.  So far, the only
complete two-loop $\order{\alpha_s}$ results exist for the four-quark
operators, where one of the loops is a fermion loop
\cite{Greub:1996tg,Buras:2001mq}. In addition, parts of two-  and
  three-loop $\order{\alpha_s^2}$ matrix element{s}, involving
gluon vacuum polarization, were found in Ref.~\cite{Bieri:2003ue}.  This last
result is very important since it automatically determines corrections to the
matrix elements of order $\alpha_s^2 \beta_0$, where~ $\beta_0 = 11 -2N_f/3$~
 is a large parameter ($N_f=5$ denotes the number of quark species).

In the present paper, we provide the first calculation of full two-loop
corrections to the matrix element of the  electromagnetic dipole operator
  $O_7$ that is responsible for the lowest-order $b\to s\gamma$ decay
rate.  We reproduce the $\alpha_s^2 \beta_0$ effect found in
Ref.~\cite{Bieri:2003ue} and also provide the $\alpha_s^2$ corrections  that
  are not enhanced by $\beta_0$.  With this result, we can check the extent
to which the $\beta_0$ effect is dominant.  The conclusion turns out to depend
very much on what renormalization scheme is used for the overall factor of
$m_b^5$ in the expression for the decay rate.

Since the result obtained here is only a partial contribution to the
future full NNLO correction, we present a number of intermediate
results and describe in detail our renormalization procedure.  We
hope that this will simplify the utilization of our result when the
remaining ingredients are known at the NNLO level.

 The paper is organized as follows. In Section~\ref{sec:calc}, we describe
  the calculation of the relevant diagrams as well as their renormalization.
  Section~\ref{sec:res} is devoted to presenting our main results and
  discussing the large-$\beta_0$ approximation. We conclude in
  Section~\ref{sec:concl}. Contributions from particular diagrams are listed
  in the Appendix.


\begin{figure}[t]
\begin{tabular}{ccccc}
\hspace{2mm}
\psfig{figure=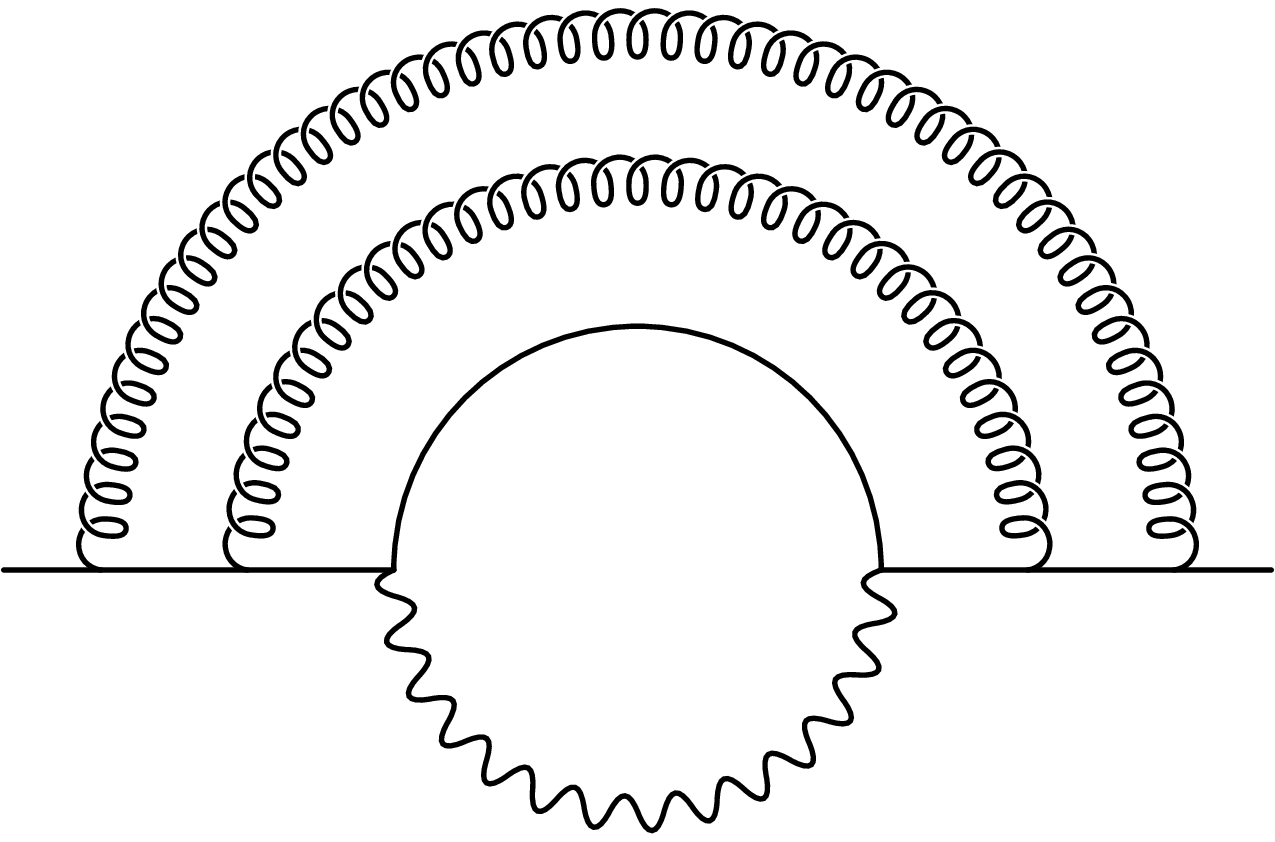,width=25mm}
\hspace{2mm}
&
\hspace{2mm}
\psfig{figure=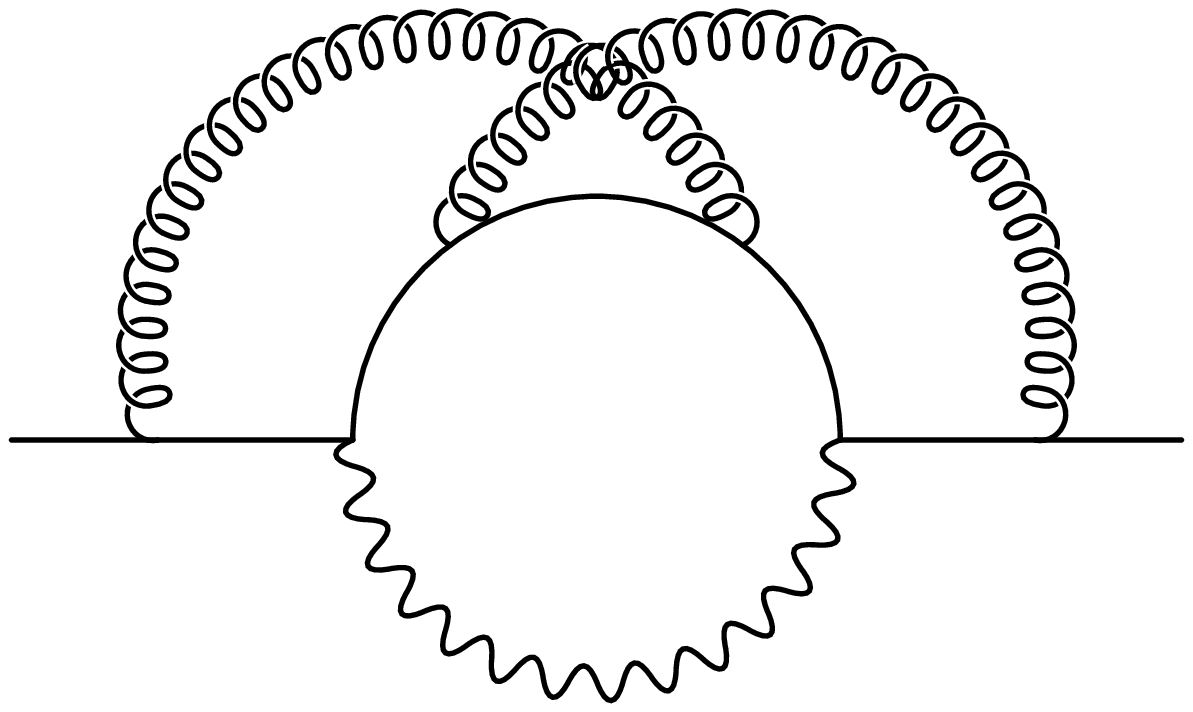,width=25mm}
\hspace{2mm}
&
\hspace{2mm}
\psfig{figure=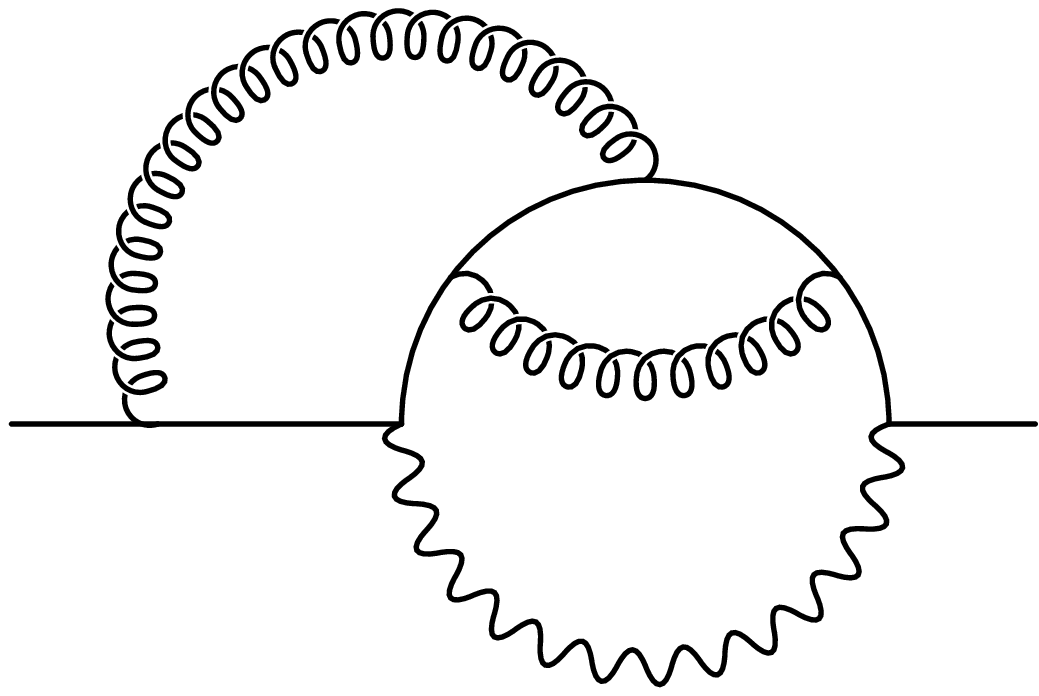,width=25mm}
\hspace{2mm}
&
\hspace{2mm}
\psfig{figure=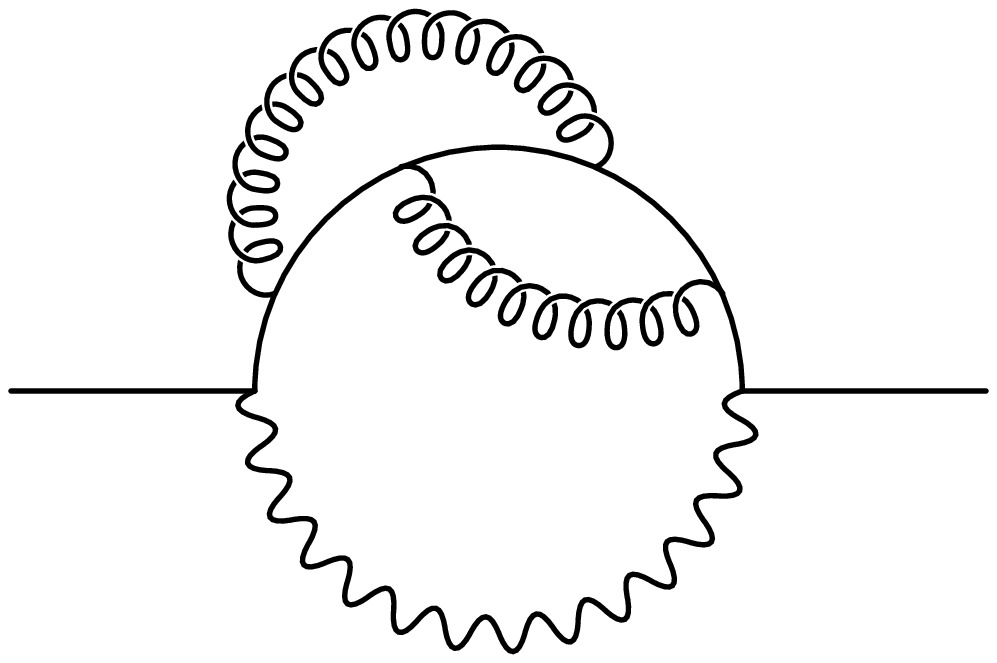,width=25mm}
\hspace{2mm}
&
\hspace{2mm}
\psfig{figure=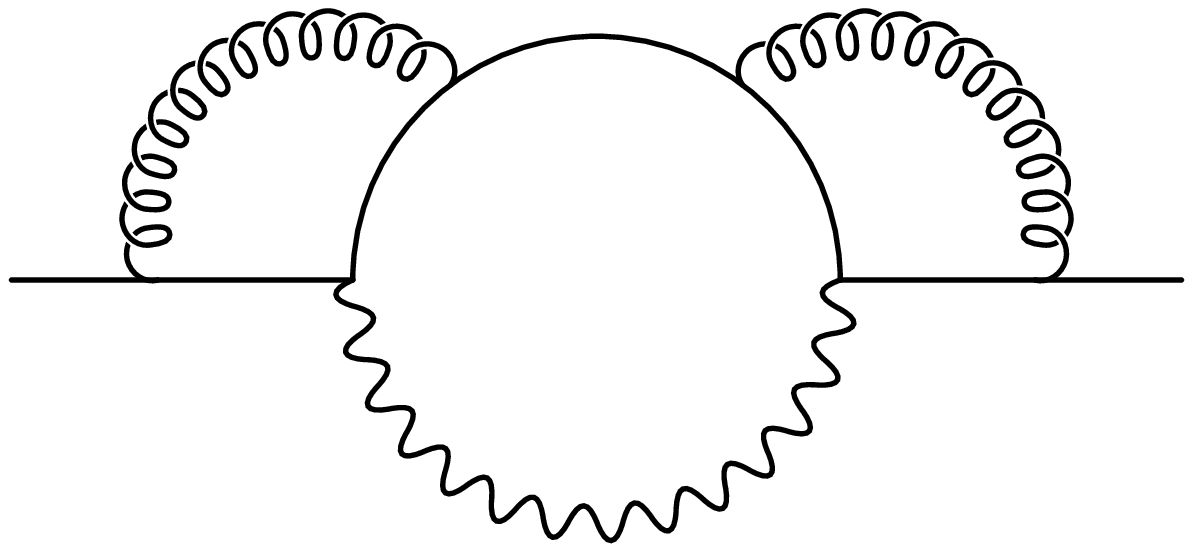,width=25mm}
\hspace{2mm}
\\
(a) & (b) & (c) & (d) & (e) \\ &&&& \\
\hspace{2mm}
\psfig{figure=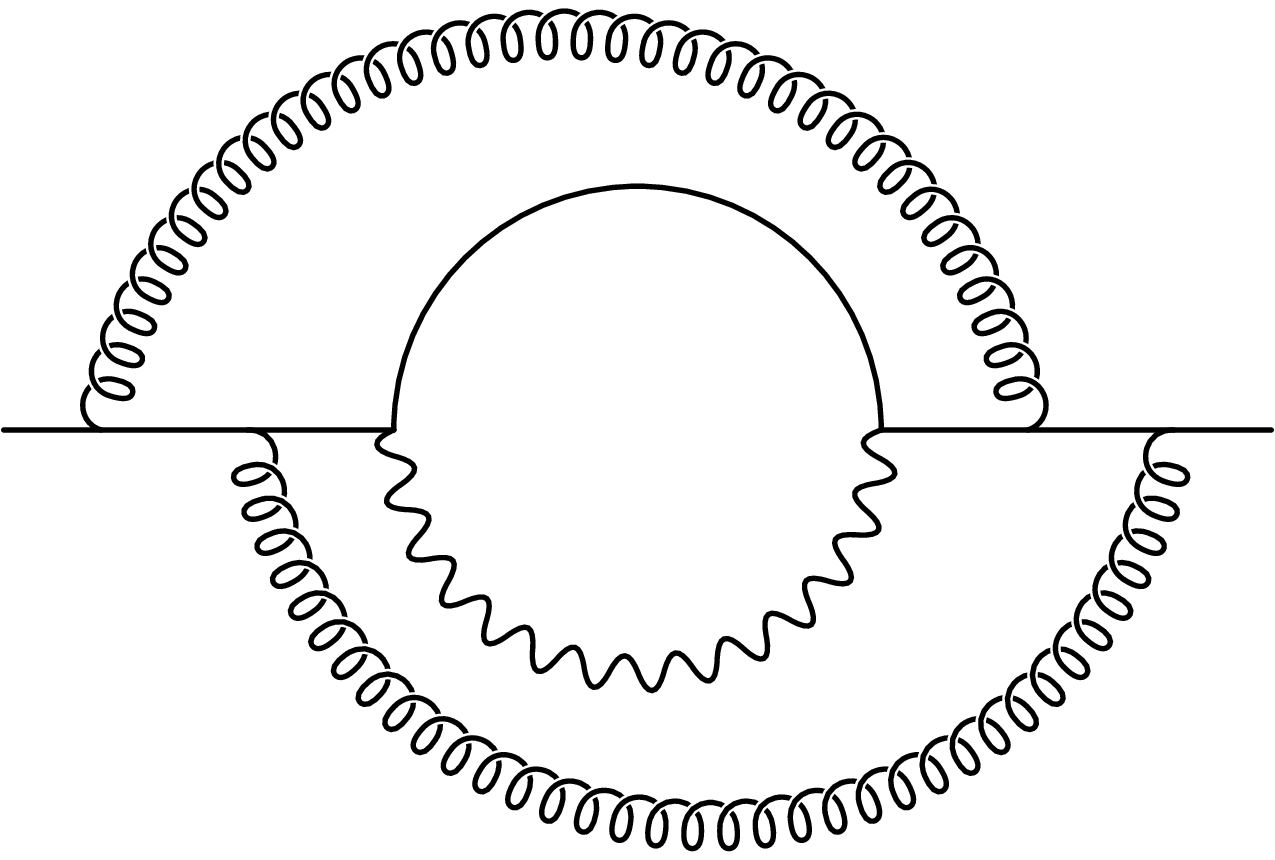,width=25mm}
\hspace{2mm}
&
\hspace{2mm}
\psfig{figure=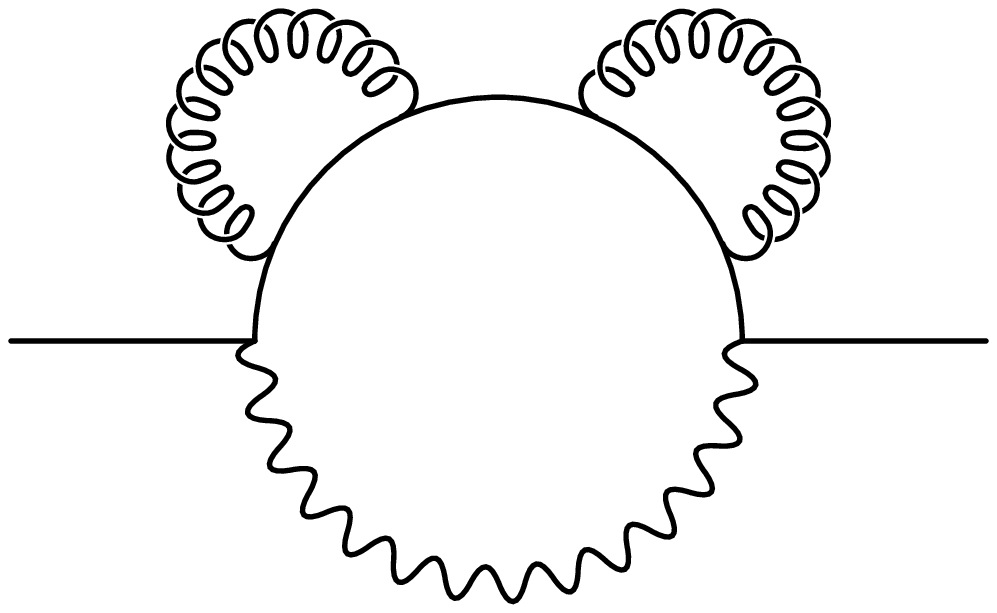,width=25mm}
\hspace{2mm}
&
\hspace{2mm}
\psfig{figure=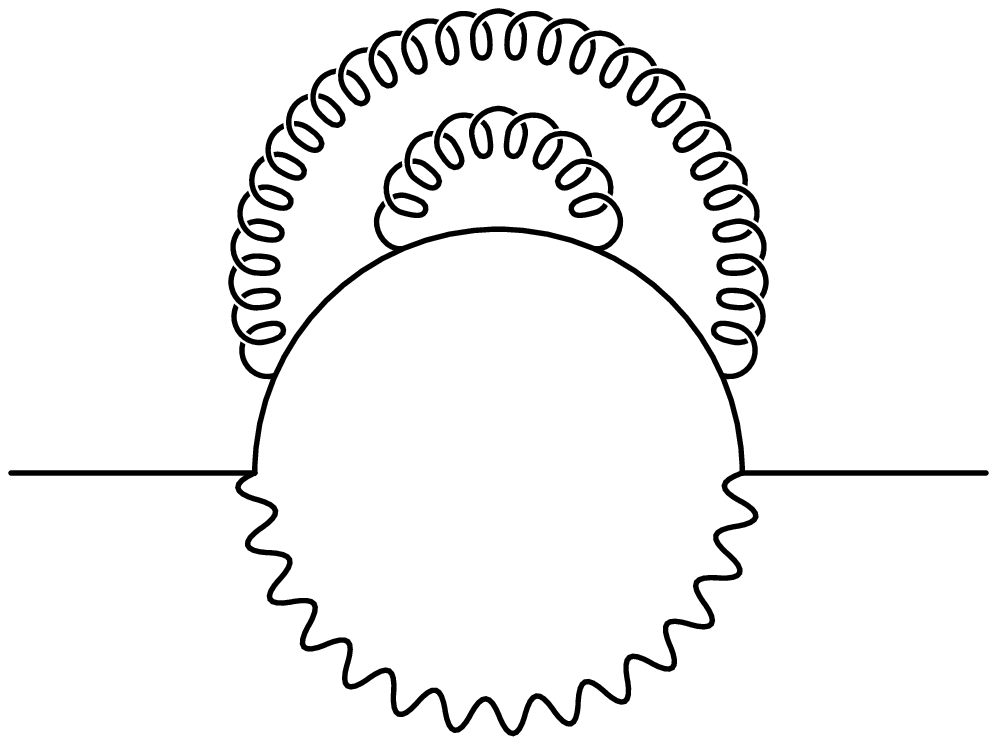,width=25mm}
\hspace{2mm}
&
\hspace{2mm}
\psfig{figure=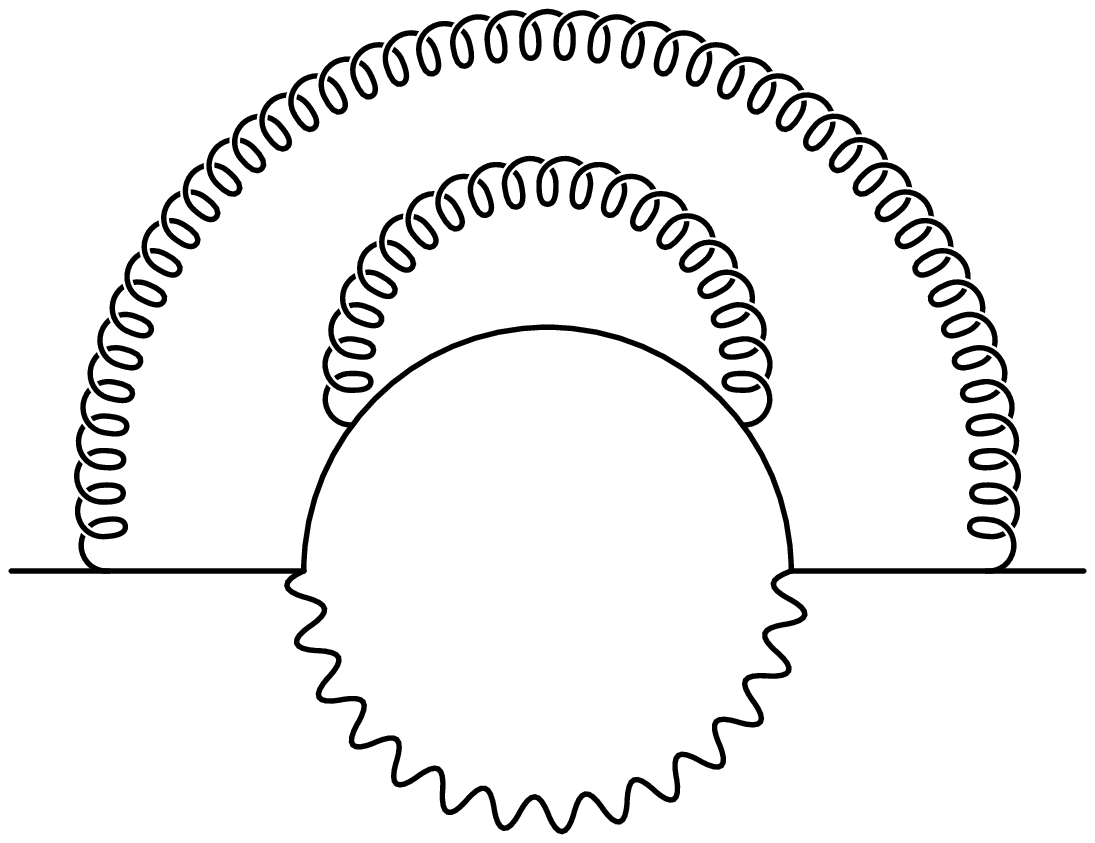,width=25mm}
\hspace{2mm}
&
\hspace{2mm}
\psfig{figure=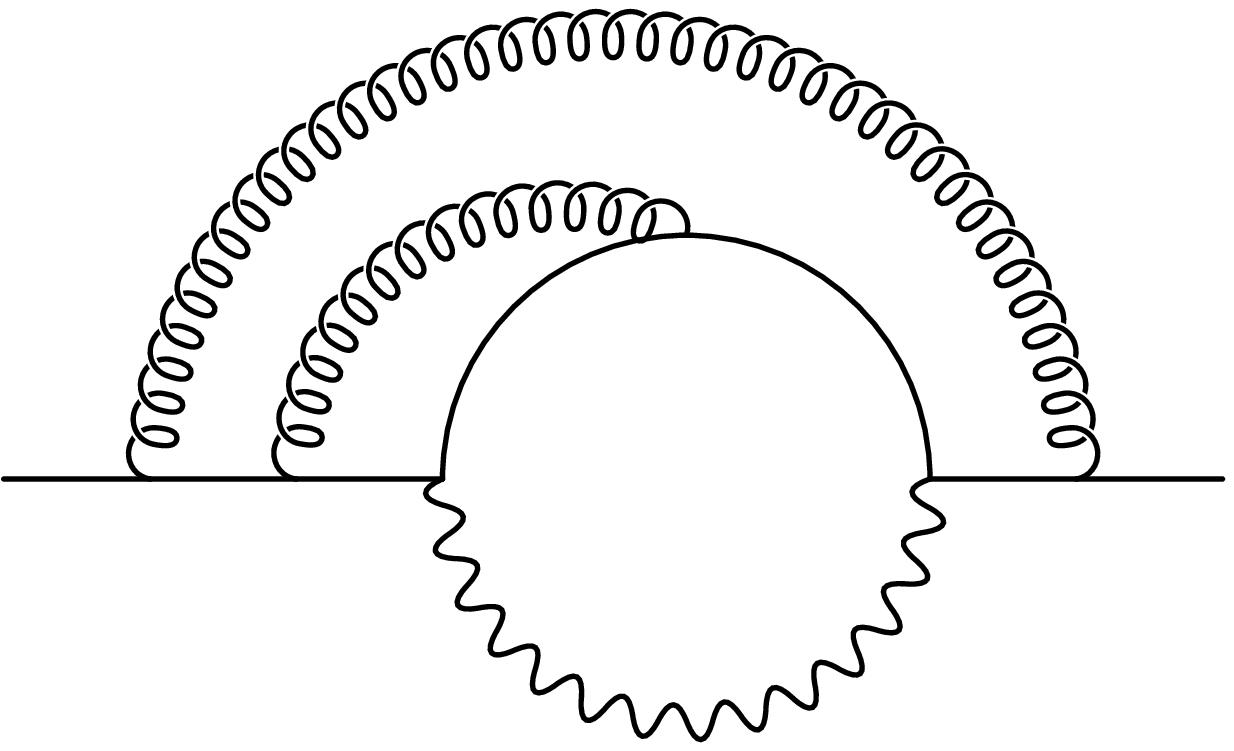,width=25mm}
\hspace{2mm}
\\
(f) & (g) & (h) & (i) & (j) \\ &&&& \\
\hspace{2mm}
\psfig{figure=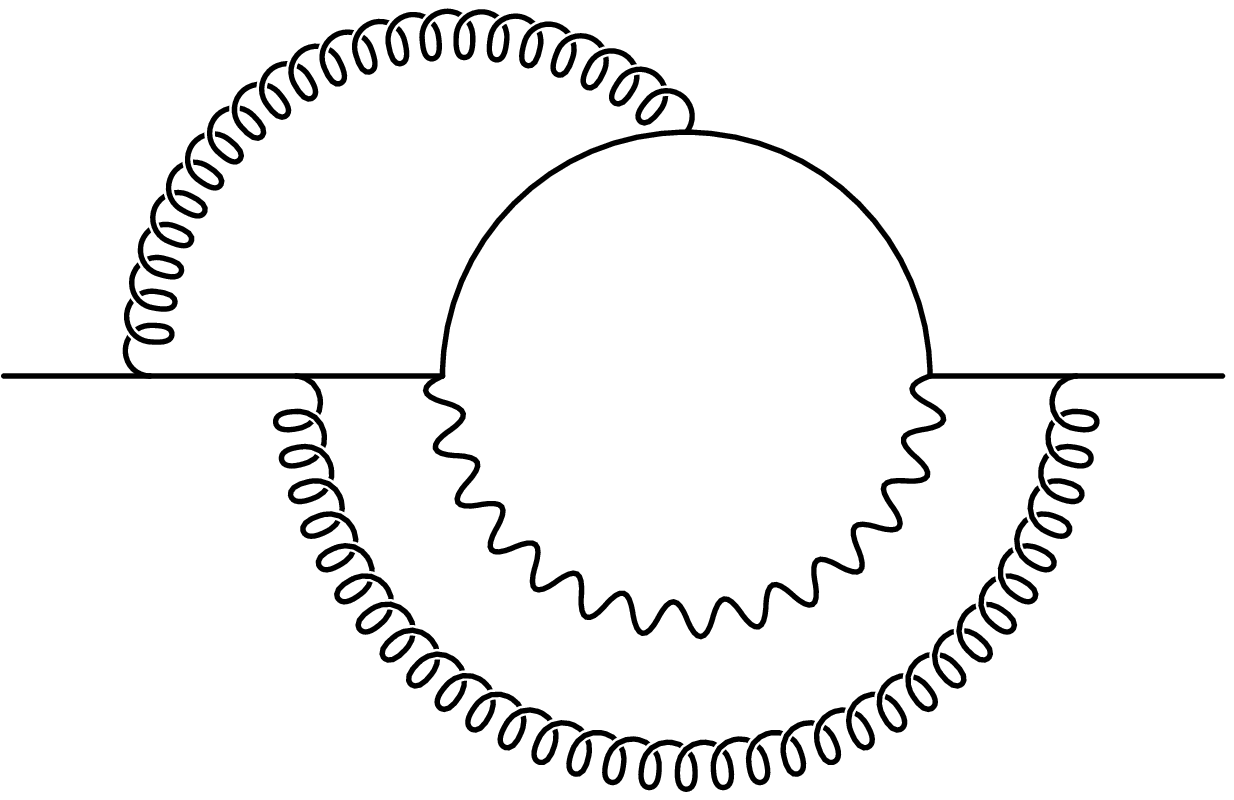,width=25mm}
\hspace{2mm}
&
\hspace{2mm}
\psfig{figure=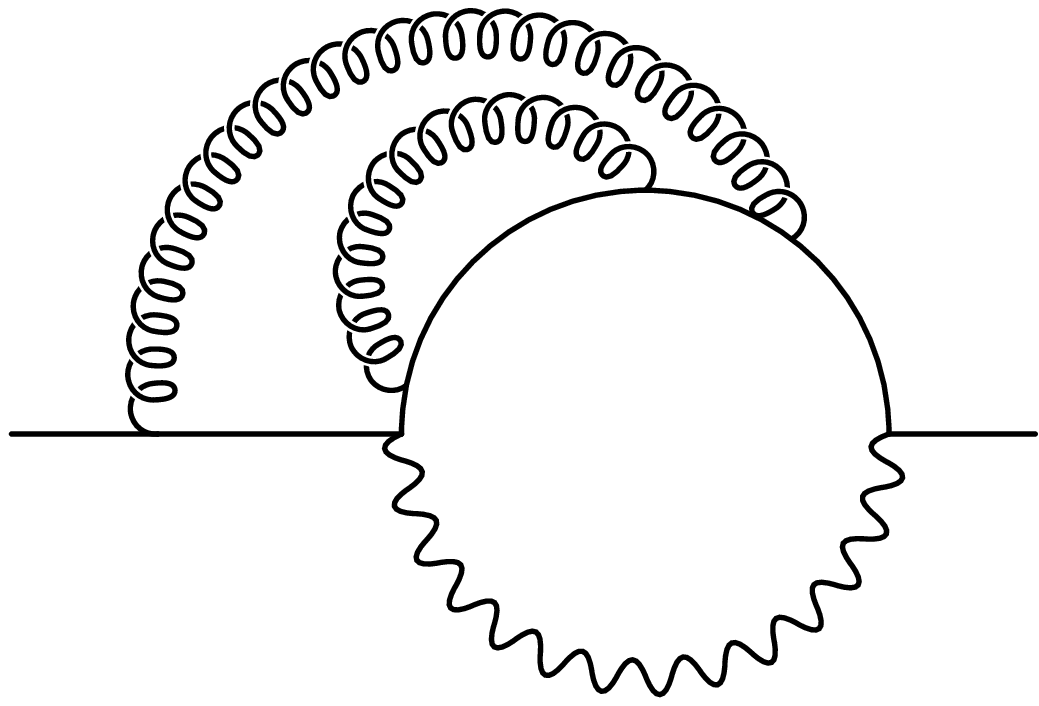,width=25mm}
\hspace{2mm}
&
\hspace{2mm}
\psfig{figure=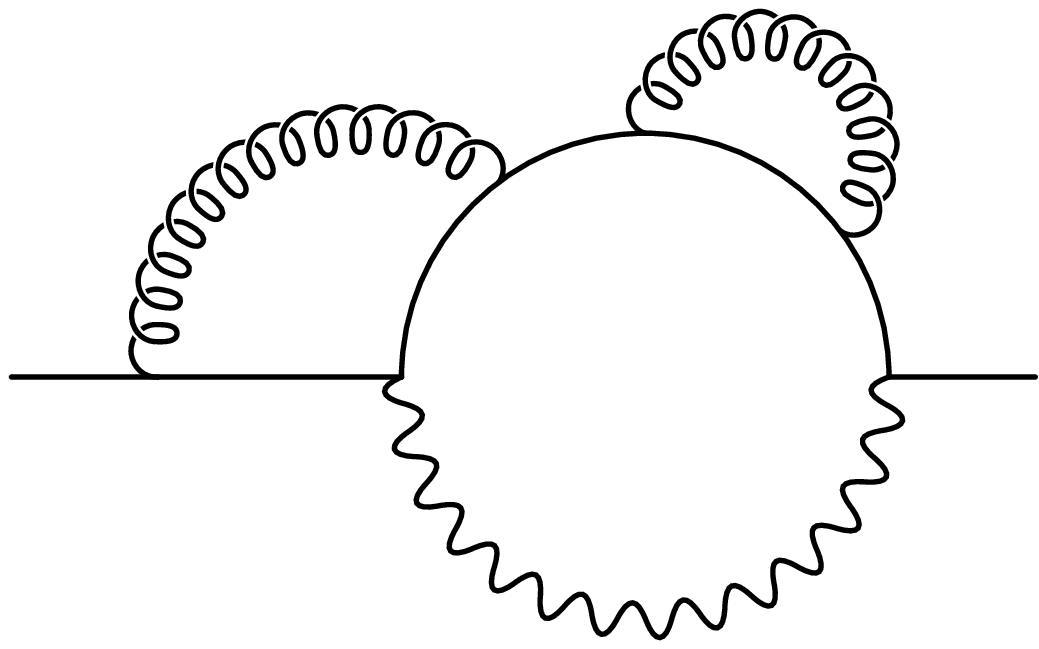,width=25mm}
\hspace{2mm}
&
\hspace{2mm}
\psfig{figure=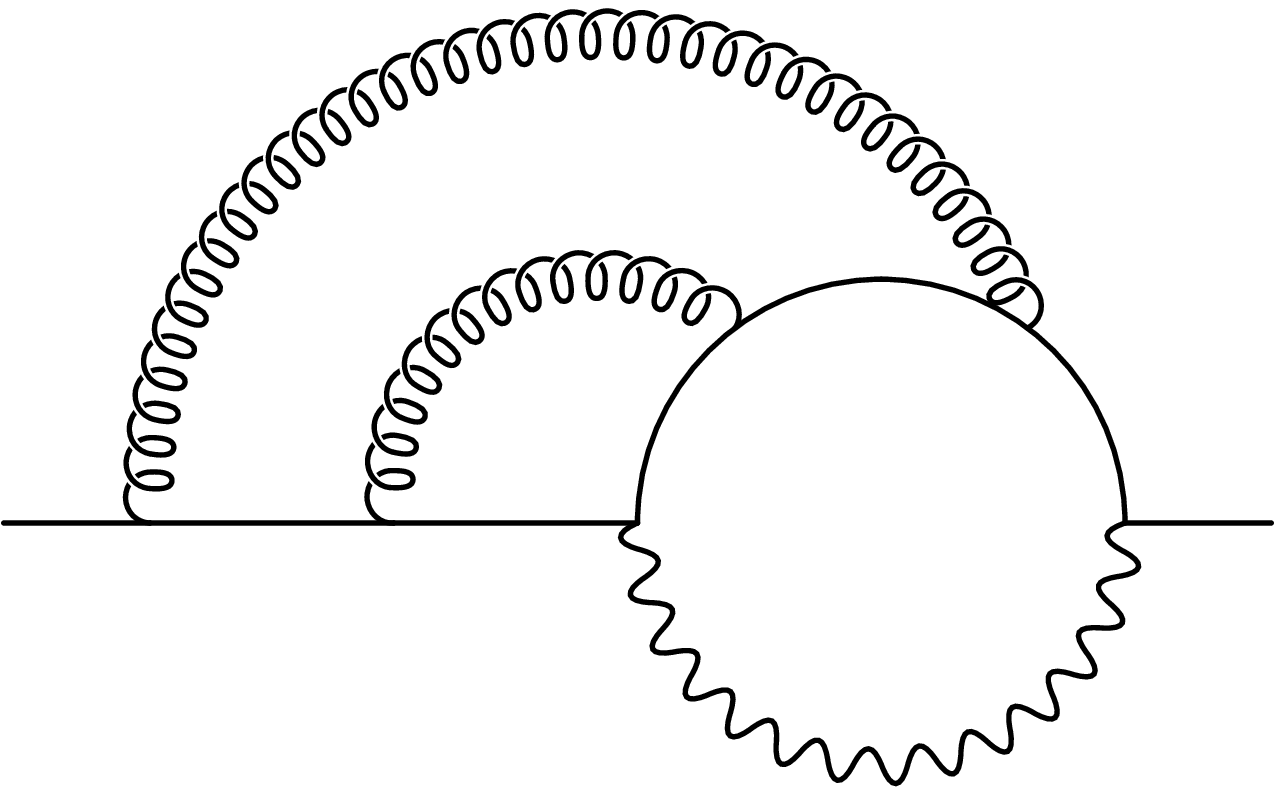,width=25mm}
\hspace{2mm}
&
\hspace{2mm}
\psfig{figure=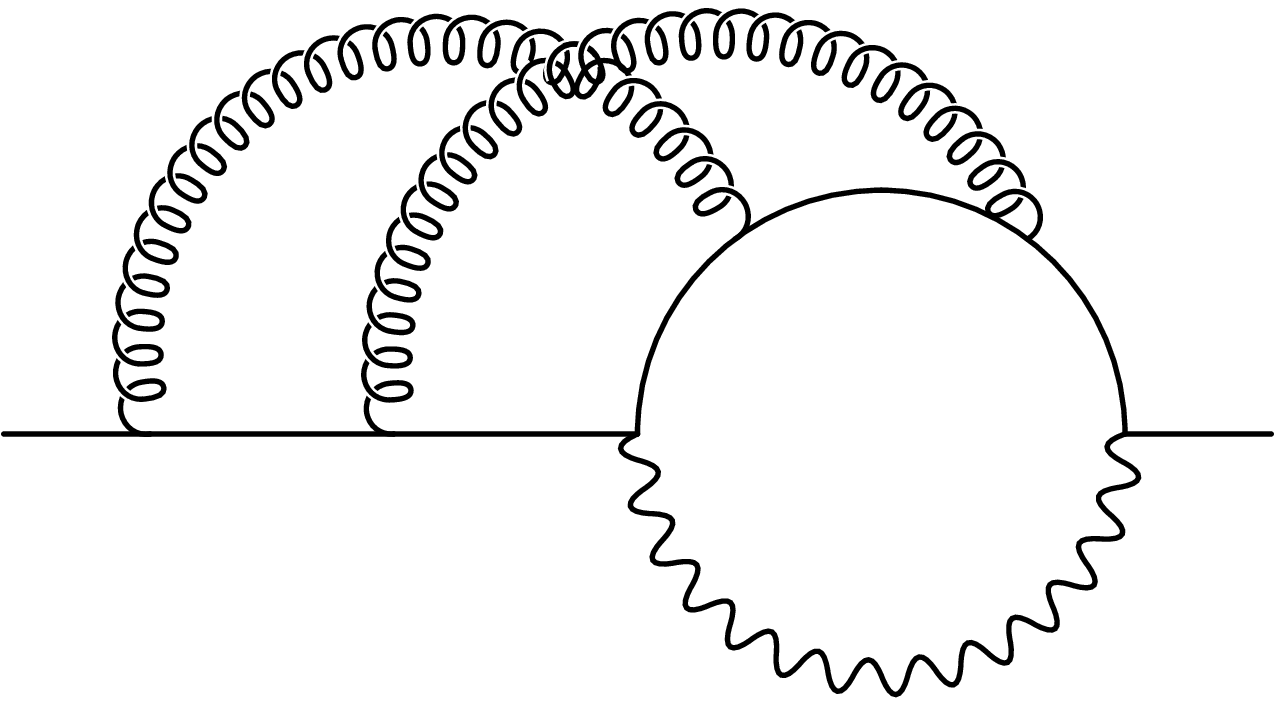,width=25mm}
\hspace{2mm}
\\
(k) & (l) & (m) & (n) & (o) \\ &&&& \\
\hspace{2mm}
\psfig{figure=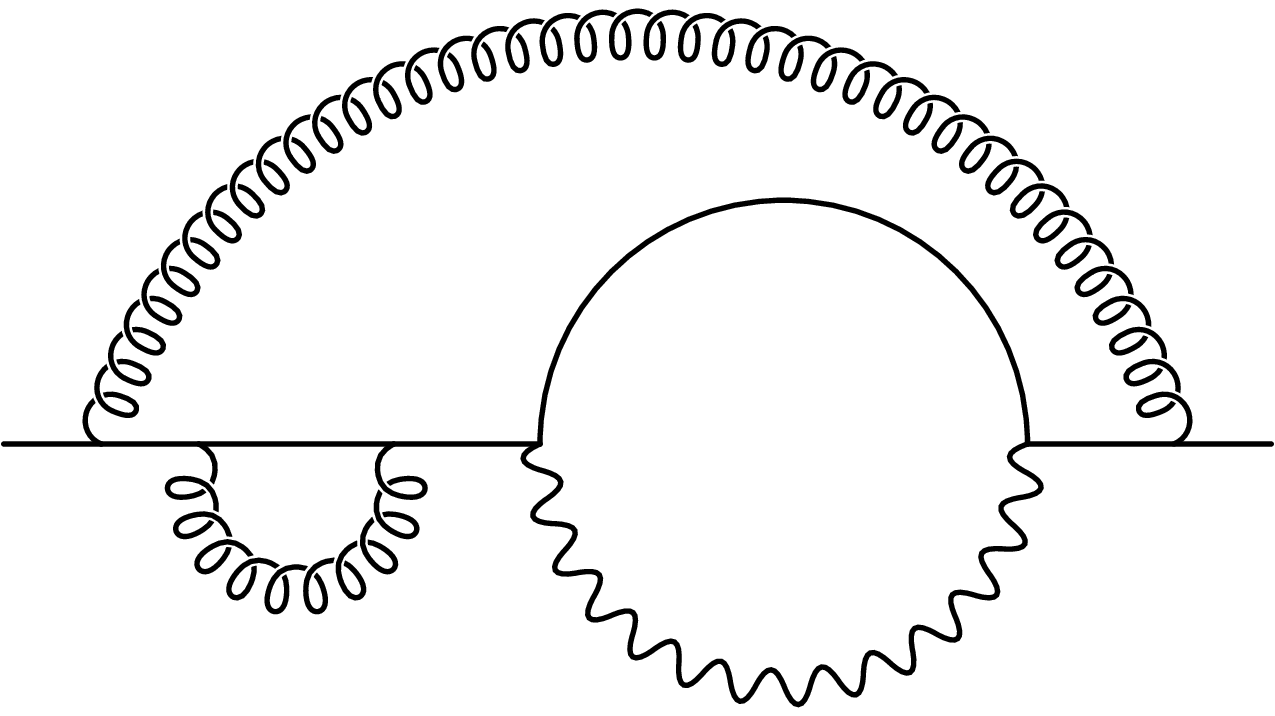,width=25mm}
\hspace{2mm}
&
\hspace{2mm}
\psfig{figure=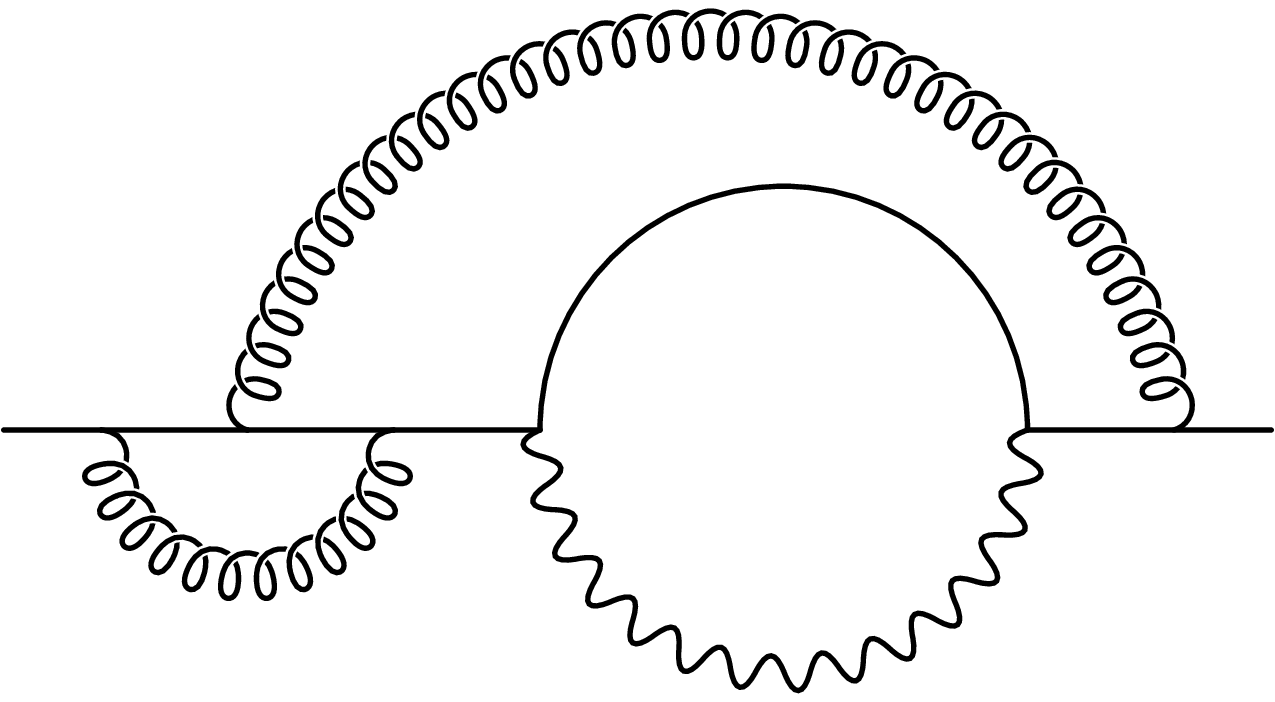,width=25mm}
\hspace{2mm}
&
\hspace{2mm}
\psfig{figure=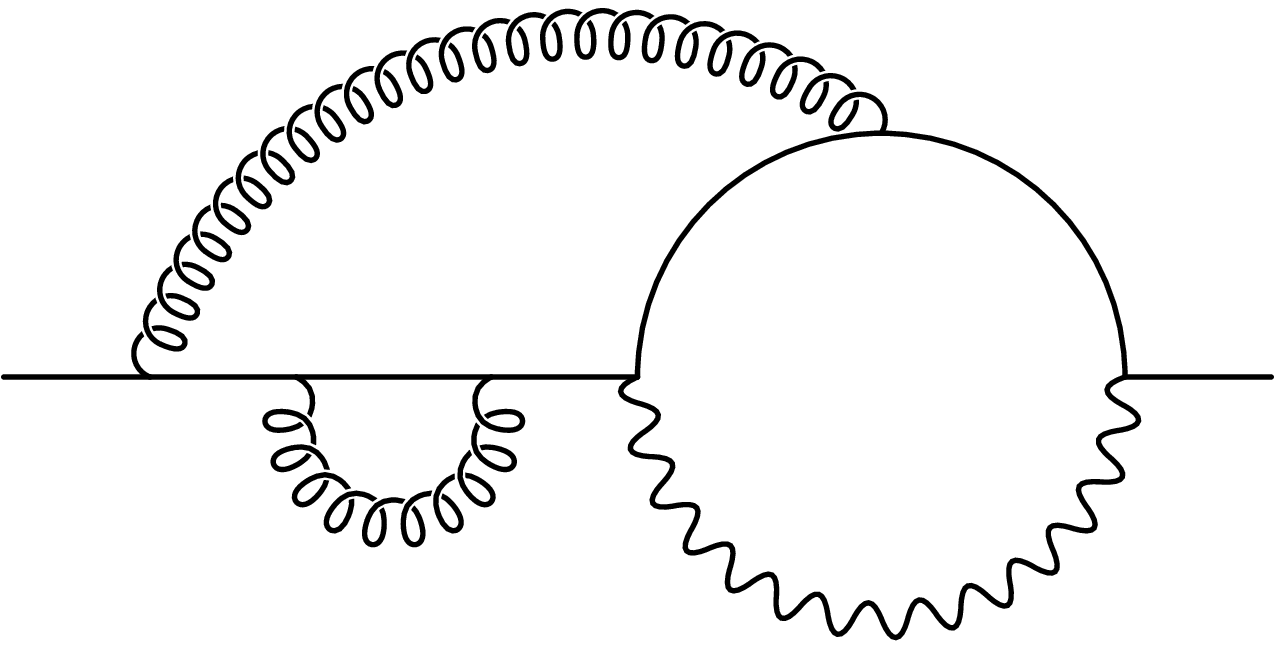,width=25mm}
\hspace{2mm}
&
\hspace{2mm}
\psfig{figure=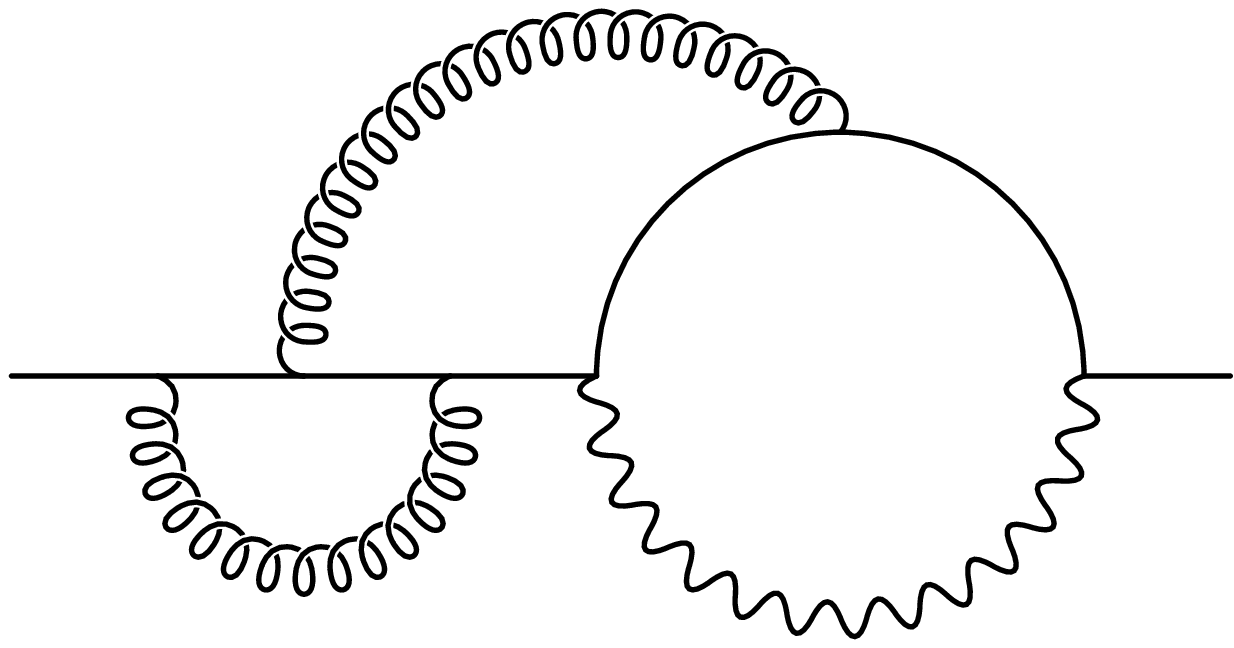,width=25mm}
\hspace{2mm}
&
\hspace{2mm}
\psfig{figure=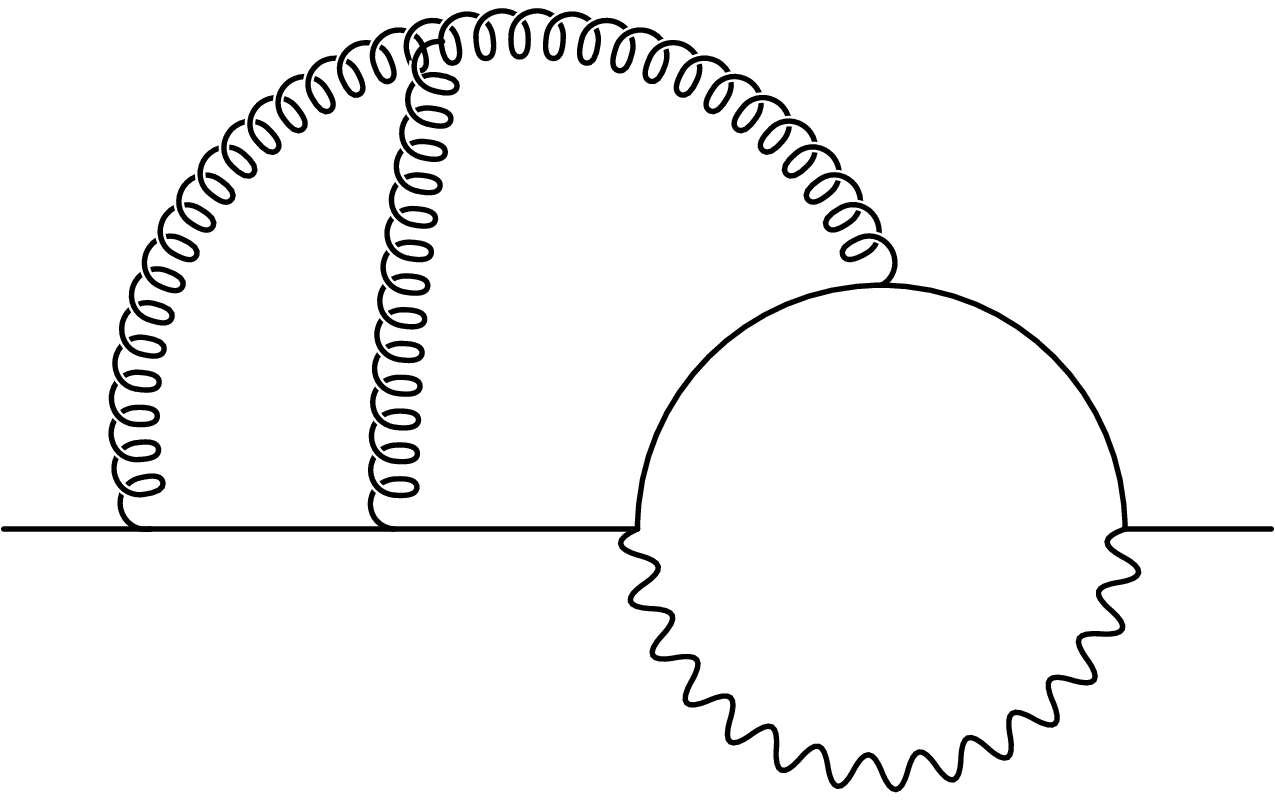,width=25mm}
\hspace{2mm}
\\
(p) & (q) & (r) & (s) & (t) \\ &&&& \\
\hspace{2mm}
\psfig{figure=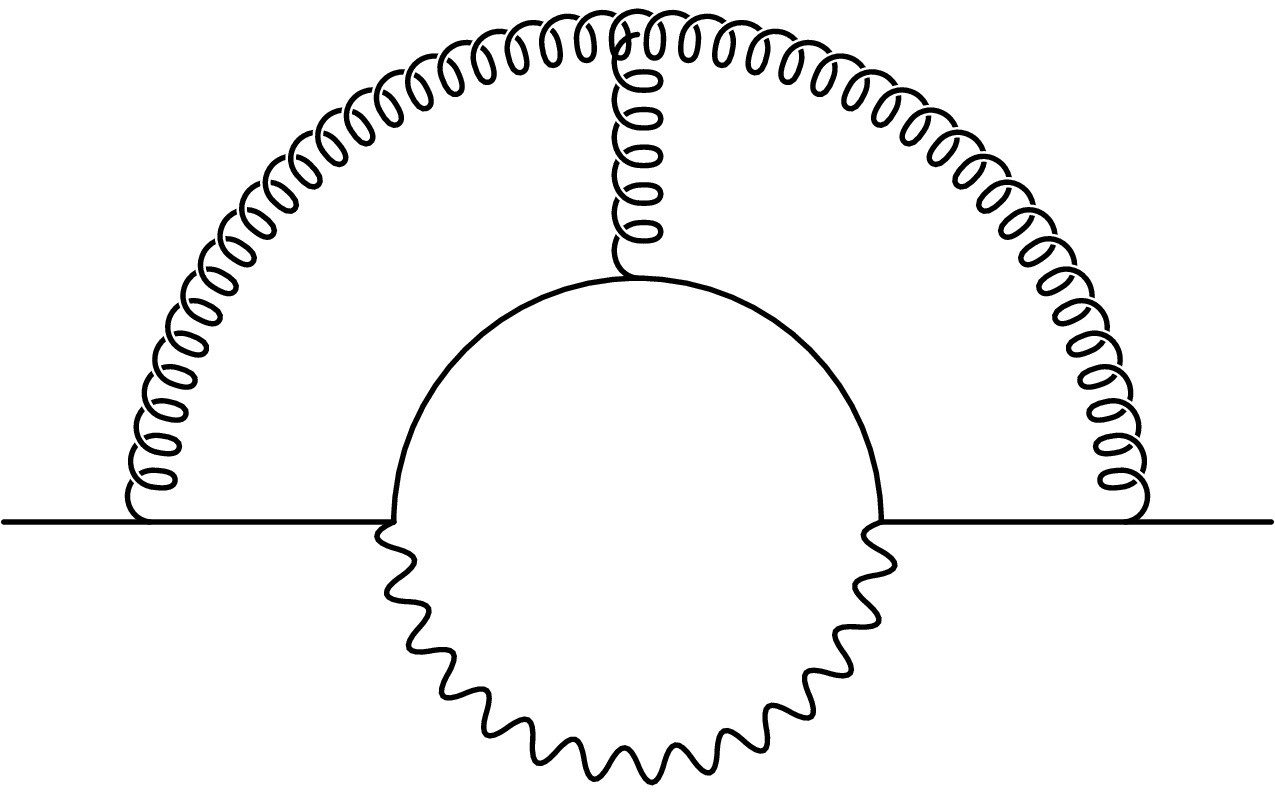,width=25mm}
\hspace{2mm}
&
\hspace{2mm}
\psfig{figure=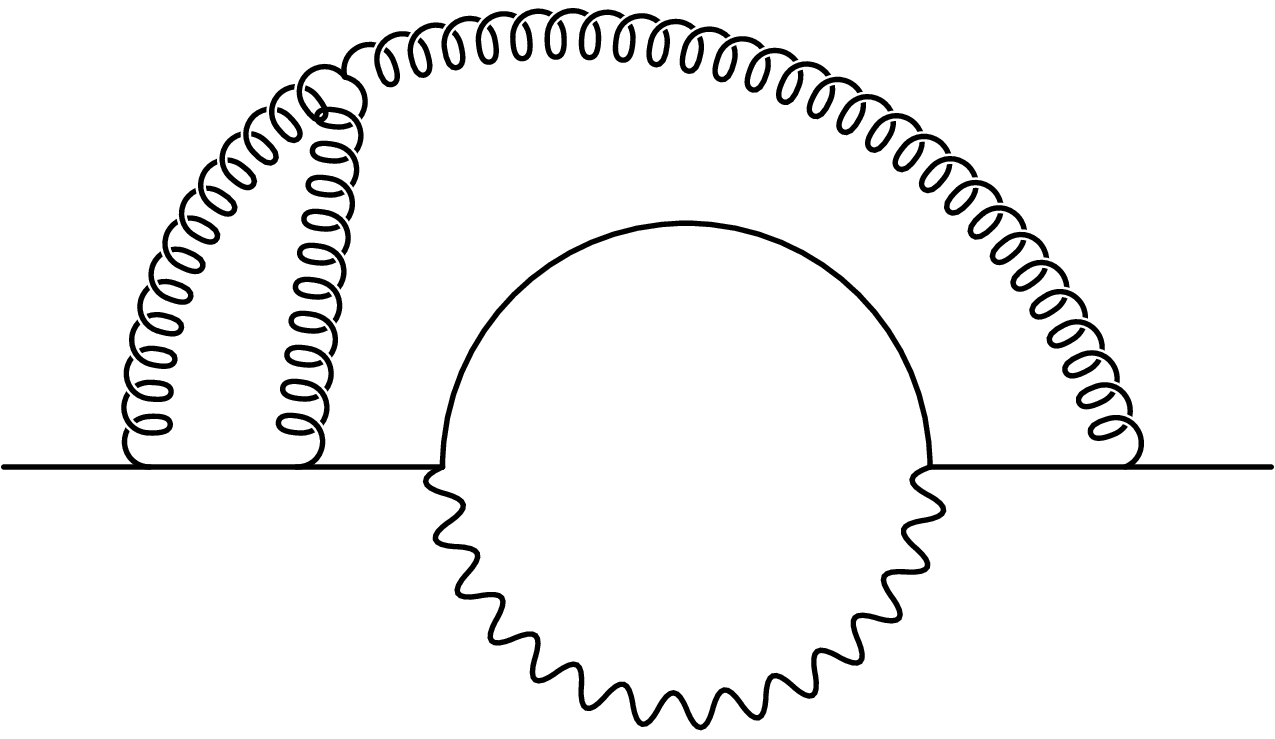,width=25mm}
\hspace{2mm}
&
\hspace{2mm}
\psfig{figure=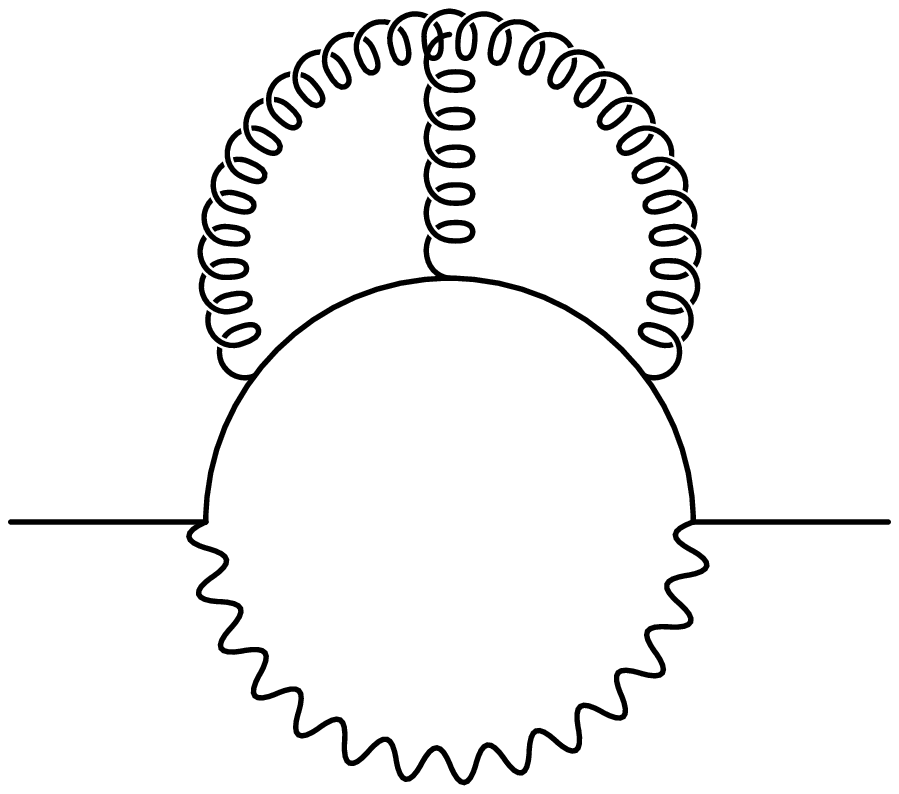,width=25mm}
\hspace{2mm}
&
\hspace{2mm}
\psfig{figure=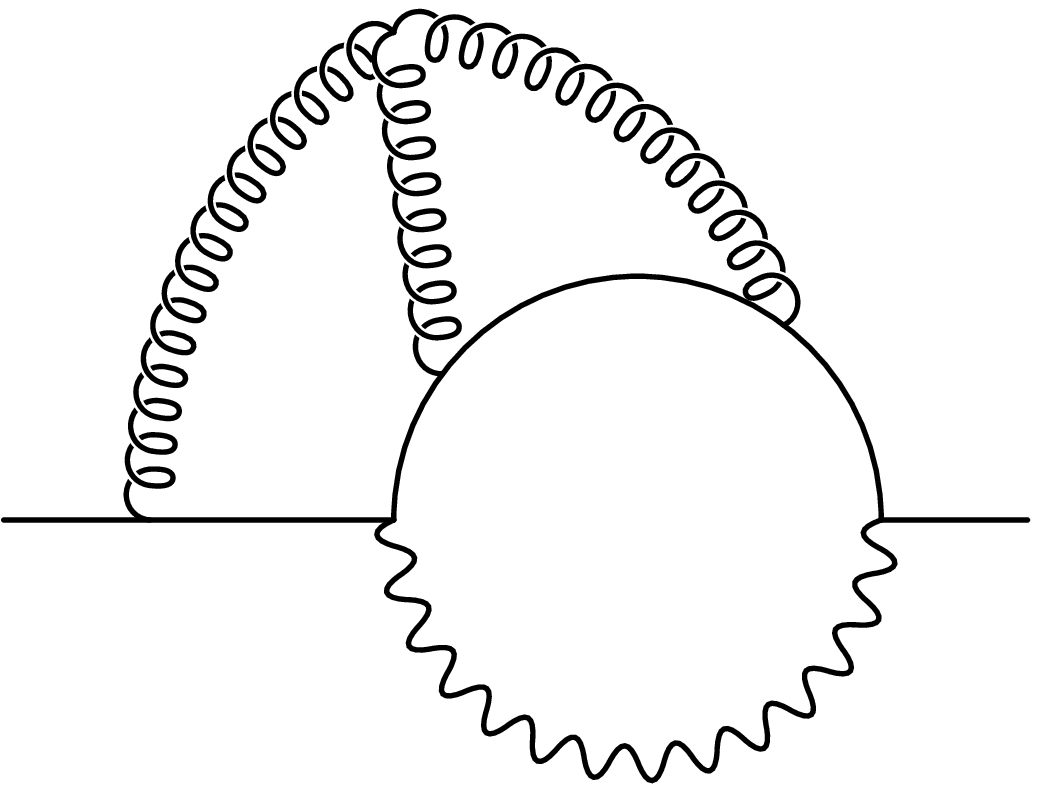,width=25mm}
\hspace{2mm}
&
\hspace{2mm}
\psfig{figure=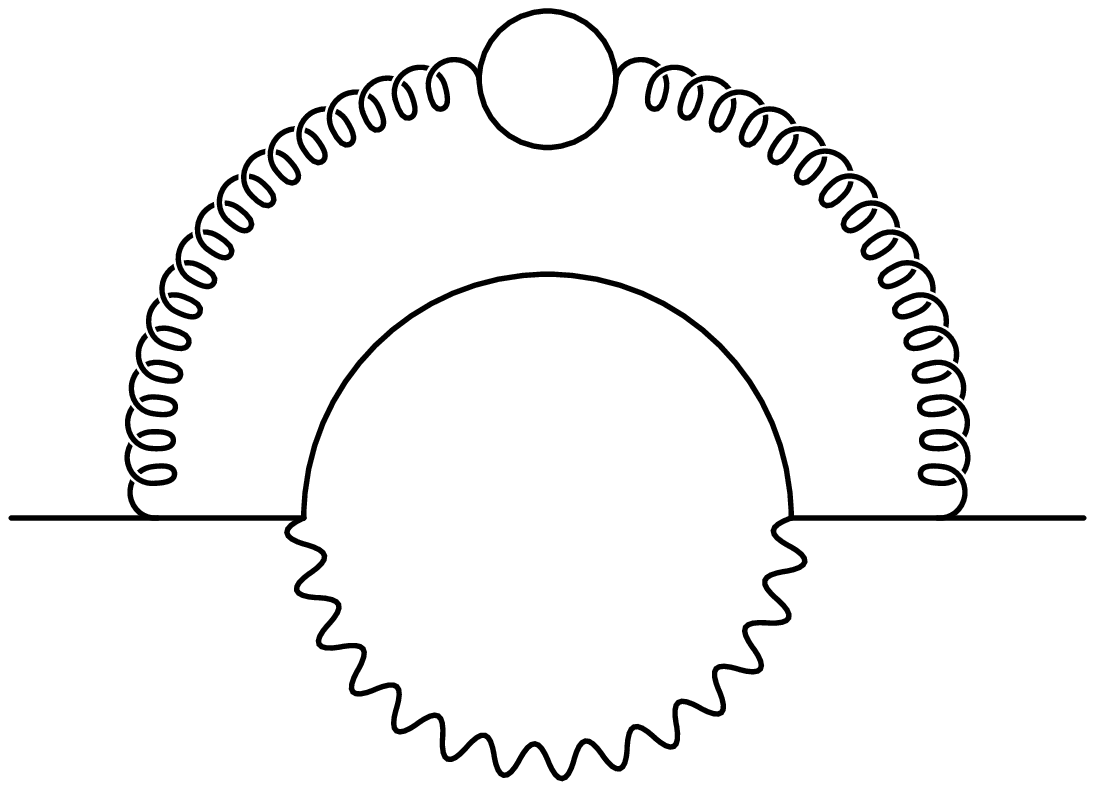,width=25mm}
\hspace{2mm}
\\
(u) & (v) & (w) & (x) & (y) \\ &&&& \\
\hspace{2mm}
\psfig{figure=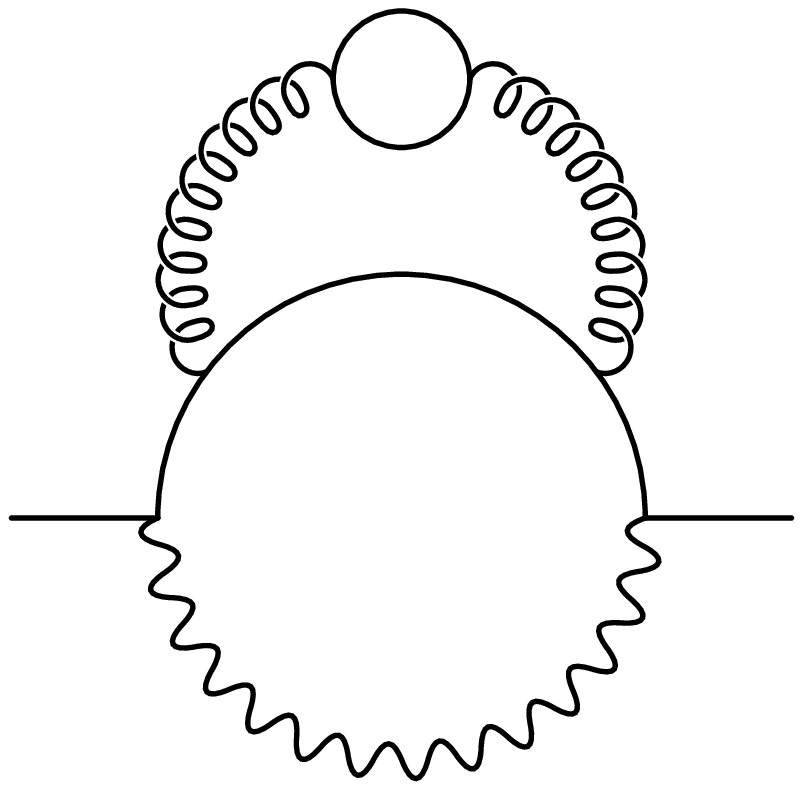,width=25mm}
\hspace{2mm}
&
\hspace{2mm}
\psfig{figure=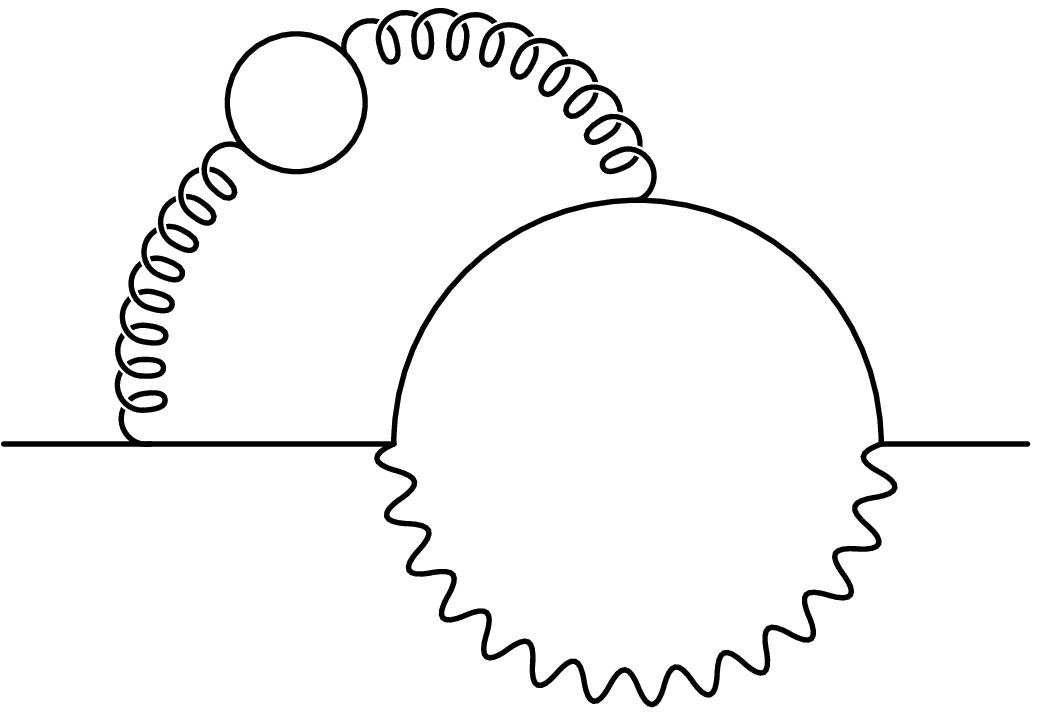,width=25mm}
\hspace{2mm}
&
\hspace{2mm}
\psfig{figure=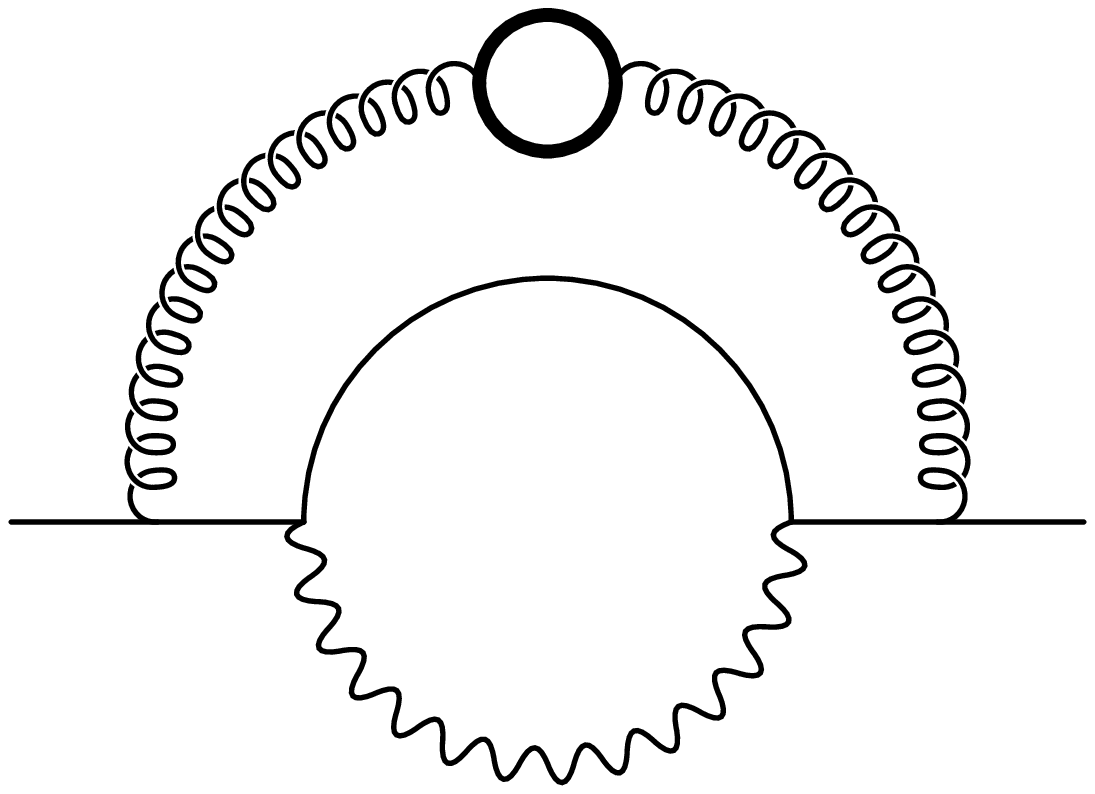,width=25mm}
\hspace{2mm}
&
\hspace{2mm}
\psfig{figure=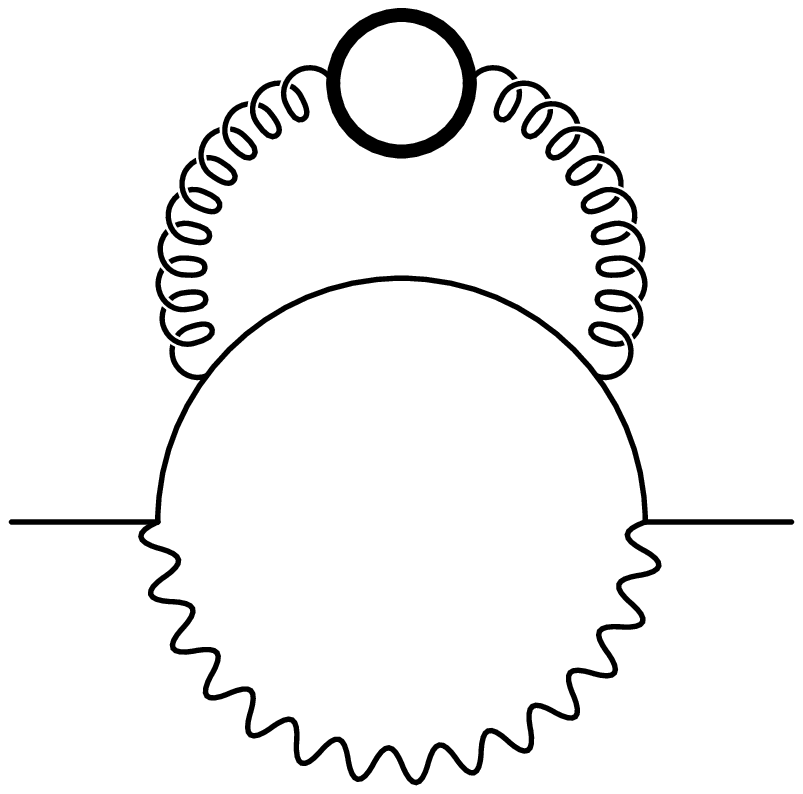,width=25mm}
\hspace{2mm}
&
\hspace{2mm}
\psfig{figure=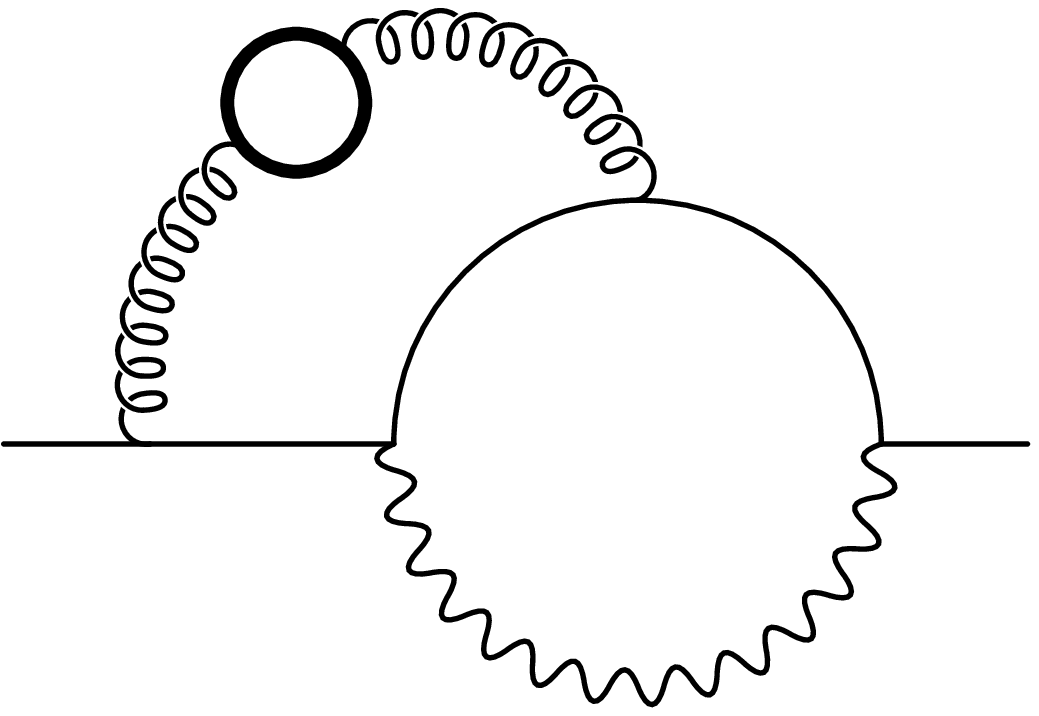,width=25mm}
\hspace{2mm}
\\
(z) & (aa) & (ab) & (ac) & (ad)
\end{tabular}
\caption{Three loop self-energy diagrams required for the
$\order{\alpha_s^2}$ computation.  Thick solid closed loops in (ab,ac,ad)
depict massive ($b$ quark) loops.
Thin solid closed loops in (y), (z), (aa) denote
massless fermions.  From the latter diagrams we find also the contributions
of gluon and ghost loops, as explained in the text.}
\label{fig:top3a}
\end{figure}

 \section{The calculation\label{sec:calc}}

The decay rate of $b$ into $s$, $\gamma$, and up to two gluons  or a light
  quark pair can be written as
\be
\Gamma ( b \rightarrow X_s^{\rm parton} \gamma )_{E_{\gamma} > E_0} =
\frac{G_F^2 \alpha_{\rm em} m_b^2(\mu) m_b^3}{32 \pi^4} |V_{tb} V_{ts}^{\ast}|^2
\sum_{i,j} C_i^{\rm eff}(\mu) \ C_j^{\rm eff}(\mu) \ G_{ij}(E_0,\mu) \, ,
\label{rate}
\ee
where $m_b$ is the pole-mass and $m_b(\mu)$ is the $\overline{\rm MS}$ running
mass  of the $b$ quark.  The effective \cite{Buras:1994xp} Wilson
  coefficients in the relevant low-energy theory are denoted by $C_i^{\rm
    eff}(\mu)$. The photon energy cutoff $E_0$ is assumed to be significantly
below the endpoint, i.e. $m_b-2E_0 \gg \Lambda_{\scriptscriptstyle\rm
  QCD}$. This is a necessary condition for the perturbative decay width
(\ref{rate}) to be a good approximation for
$\Gamma ( \bar{B} \rightarrow X_s \gamma )_{E_{\gamma} > E_0}$.

In this paper{,}  we focus on the contribution of the operator
\be
O_7 = {e m_b(\mu) \over 16\pi^2} \,  \bar s_L \sigma^{\mu \nu}b_R \, F_{\mu \nu}
\ee
to the decay rate.  More specifically,  we calculate
\be \label{g77}
G_{77}(0,\mu) = 1 + \left( \frac{\alpha_s(\mu)}{4 \pi} \right) X_1
                  + \left( \frac{\alpha_s}{4 \pi} \right)^2 X_2 + \order{ \alpha_s^3} \ .
\ee
 Our result for $G_{77}(0,\mu)$ can be combined with the very recent
  findings of Ref.~\cite{Melnikov:2005bx} to obtain $G_{77}(E_0,\mu)$ at the
  NNLO for any value of $E_0$ that is sufficiently far from the endpoint.

  As far as the coefficient $X_1$ is concerned, we confirm the well-known
  result of Ref.~\cite{Ali:1990tj}. The  NNLO  correction $X_2$ can be
subdivided into color structures,
\begin{equation}
X_2 = C_F\left( T_R N_L X_L + T_R N_H X_H + C_F X_A + C_A X_{NA}
\right) \ ,
\label{X2pieces}
\end{equation}
where $C_F=4/3$, $C_A=3$, and $T_R=1/2$ are the SU(3) color factors,
  while $N_L$ and $N_H$ denote the number of light $(m_q=0)$ and heavy
$(m_q=m_b)$ quark species  ($N_H+N_L=N_f$).

Let us now briefly outline  the method  applied to the calculation of
$X_2$. We use the optical theorem to map all  the  virtual corrections
and real radiation contributions onto a system of self-energy diagrams as
described in Ref.~\cite{Blokland:2004ye} in the context of the decays $t\to
X_b W$ and $b\to X_u l\bar{\nu}$. The topologies which have to be taken into
account turn out to be identical to those considered there for the top quark
decay.

We consider only gluons of virtualities  of order $m_b$. In the effective
theory, this is the only mass scale.  In other words, we do not consider
hard gluons that would resolve the structure of the effective vertex
 $\bar{s}b\gamma$ since their effects are accounted for in the Wilson
coefficients  $C_i^{\rm eff}( \mu )$.  Diagrams needed for the present
calculation are analogous to those of the hard asymptotic region for the
top quark decay, studied recently in Ref.~\cite{Blokland:2004ye,Blokland:2005vq}.

Diagrams contributing to $X_2$ are presented in Fig.~\ref{fig:top3a}.  All
  the particles except for the $b$ quarks are treated as massless. Both the UV
  and the IR divergences are regulated dimensionally in $D=4-2\ep$ dimensions.
   The results  for $\mu=m_b$ are collected in Table~I in the
Appendix along with the relevant color factors. They have been computed in a
general covariant gauge.  The Feynman gauge results can be obtained by setting
$\xi = 0$.  Diagrams that are not symmetric under left-right reflection
  are already multiplied by 2. The quantities $\ln 4\pi$ and $\gamma$ that
  account for the difference between the ${\rm MS}$ and $\overline{\rm MS}$
  schemes are omitted in Table~I and in the remainder of this section.

Diagrams with closed gluon or ghost loops are not shown explicitly in Fig.~\ref{fig:top3a},
since their contribution can be found from the total contribution of the light fermions.
To this end, after computing the light-fermion diagrams,
we replace (see for example Ref.~\cite{Muta:1987mz})
\begin{eqnarray} \label{muta.replacement}
T_R N_L \to T_R N_L -C_A \left[{5 \over 4}-{3 \over 8}\xi
     +\ep \left({1 \over 2} +{11 \over 8}\xi+ {3 \over 16}\xi^2 \right)
     +\ep^2\left({1 \over 2} +{3 \over 8}\xi +{1 \over 16}\xi^2\right)  +\order{\ep^3} \right].
\end{eqnarray}

  Let $B_3$ denote the sum of all the three-loop diagrams from Table~I, after
  performing the above replacement. Our final result for $G_{77}(0,\mu)$ is
  found according to the following formula
\ba \label{all.renor}
G_{77}(0,\mu) &=& \frac{1}{16}
\left( Z_m^{\overline{\rm MS}} Z_{77}^{\overline{\rm MS}} \right)^2 Z_{\psi}^{\rm OS} \left[ B_1 +
\frac{\alpha_s(\mu)}{4\pi} Z_{\alpha}^{\overline{\rm MS}} \frac{\mu^{2\ep}\kappa}{m_b^{2\ep}}
\left( B_2 + (Z_m^{\rm OS}-1) B_2^m \right) + \left(\frac{\alpha_s}{4\pi}\right)^2
\frac{\mu^{4\ep}\kappa^2}{m_b^{4\ep}} B_3 \right] + \order{\alpha_s^3},
\ea
where
\ba
B_1 &=& 16 + 16\ep + \left( 32 - {16 \over 3}\pi^2 \right)\ep^2 +\order{\ep^3},\\
B_2 &=& C_F \left[ \frac{16}{\ep} +\frac{496}{3} -\frac{64}{3}\pi^2
+ \ep \left( 848 - 80\pi^2 - 256\zeta_3 \right) +\order{\ep^2} \right]
\ea
stand for the one-loop (Born) diagram and the sum of two-loop diagrams, respectively,
while $\kappa = 1 + \ep^2 \pi^2/12$.
The  $\overline{\rm MS}$ renormalization constant for the operator vertex
\begin{equation}
 Z_m^{\overline{\rm MS}} Z_{77}^{\overline{\rm MS}}  =
1+ {\alpha_s(\mu) \over 4\pi} {C_F \over \ep}
 + \left({\alpha_s\over 4\pi} \right)^2 {C_F \over \ep}\left[
{257 \over 36} C_A - {19\over 4} C_F - {13 \over 9} T_R N_f +
{1\over \ep} \left( {1\over 2} C_F - {11\over 6} C_A +{2\over 3} T_R
N_f\right)\right]  + \order{\alpha_s^3}
\end{equation}
is found from the anomalous dimensions published in Ref.~\cite{Misiak:1994zw}.
 The on-shell renormalization constant of the quark field can be
  written as \cite{Broadhurst:1991fy}
\be
Z_{\psi}^{\rm OS} = 1 + \left( \frac{\alpha_s(\mu)}{4 \pi} \right) \kappa P_1
                  + \left( \frac{\alpha_s}{4 \pi} \right)^2 \kappa^2 P_2 + \order{\alpha_s^3},
\ee
where
\ba
P_1 &=& C_F \left[ - \frac{3}{\ep}
-4 + 6 \ln{m_b\over \mu} + \ep \left( -8 + 8 \ln{m_b\over \mu}
- 6 \ln^2{m_b\over \mu} \right) + \order{\ep^2}\right],\\
P_2 &=& C_F\left( T_R N_L P_L + T_R N_H P_H + C_F P_A + C_A P_{NA} \right),
\ea
and
\ba
P_L &=& - \frac{2}{\ep^2} + \frac{11}{3\ep} + \frac{113}{6} + \frac{5}{3}\pi^2
- \frac{76}{3} \ln{m_b\over \mu} + 8 \ln^2{m_b\over \mu} +\order{\ep},\\
P_H &=& \frac{1}{\ep} - \frac{8}{\ep} \ln{m_b\over \mu} + \frac{947}{18} -
5 \pi^2 - \frac{44}{3} \ln{m_b\over \mu}
+ 24 \ln^2{m_b\over \mu} +\order{\ep},\\
P_{NA} &=& \frac{11}{2 \ep^2} - \frac{127}{12 \ep} - \frac{1705}{24}
+ \frac{49}{12} \pi^2 - 8 \pi^2 \ln 2 + 12 \zeta_3 + \frac{215}{3} \ln{m_b\over
\mu} - 22 \ln^2{m_b\over
  \mu} +\order{\ep},\\
P_A &=& \frac{9}{2 \ep^2} + \frac{1}{\ep} \left( \frac{51}{4} - 18
  \ln{m_b\over \mu} \right) + \frac{433}{8} - 13 \pi^2 + 16 \pi^2 \ln 2  - 24
\zeta_3 - 51 \ln{m_b\over \mu} + 36 \ln^2{m_b\over \mu}
+\order{\ep}. \ea
Mass renormalization in the $b$ quark propagators is accounted for by squaring
these propagators in the two-loop diagrams, which turns $B_2$ into
\be
B_2^m = C_F \left[ \frac{1}{\ep} \left( 96 - 48 \xi \right) + 656 - 16 \xi -
  64 \pi^2 + \ep \left( 2344 - 256 \xi - 160 \pi^2 + 16 \pi^2 \xi - 768
    \zeta_3 \right) \right].
\ee
In the expression (\ref{all.renor}) for $G_{77}$, the above quantity gets
multiplied by $Z_m^{\rm OS} - 1 = \frac{\alpha_s}{4\pi}\kappa P_1 +
\order{\alpha_s^2}$.  For completeness, the one-loop gauge coupling
renormalization constant should also be mentioned
\be \label{zalpha}
Z_{\alpha}^{\overline{\rm MS}} = 1 + \left( \frac{\alpha_s}{4 \pi \ep} \right)
\left( {4\over 3} T_R N_f - {11\over 3} C_A \right) + \order{\alpha_s^2}.
\ee

 \section{ Results\label{sec:res}}

 Our final results for the contributions to $G_{77}(0,\mu)$ read
\begin{eqnarray}
     X_1&=&  C_F \left( {16 \over 3} + 4 \ln{m_b\over \mu} - {4 \over 3} \pi^2  \right) \ ,
\nonumber \\
     X_L&=&   - {251 \over 27} +\left(- {32 \over 9} \pi^2 + {8 \over 3}\right)
         \ln{m_b\over \mu} + {16 \over 3} \ln^2{m_b\over \mu}
         + 16 \zeta_3 + {128 \over 27} \pi^2   \ ,
\nonumber \\
X_H &=&   {7126 \over 81} +\left(- {32 \over 9} \pi^2 + {8 \over 3}\right)
         \ln{m_b\over \mu}
         + {16 \over 3} \ln^2{m_b\over \mu} - {16 \over 3} \zeta_3
         - {232 \over 27} \pi^2    \ ,
\nonumber \\
X_{NA} &=&  - {1333 \over 216} +\left( {88 \over 9} \pi^2 + 18\right)
         \ln{m_b\over \mu} - {44 \over 3} \ln^2{m_b\over \mu}
         - {47 \over 6} \zeta_3 - 27 \pi^2 \ln 2 + {119 \over 108}
         \pi^2 + {43 \over 90} \pi^4   \ ,
\nonumber \\
X_A &=&   {2825 \over 18} - \left({16 \over 3}  \pi^2+ {50 \over 3}\right) \ln{m_b\over \mu}
          + 8 \ln^2{m_b\over \mu} - {217 \over 3} \zeta_3 + 54 \pi^2 \ln 2
           - {319 \over 6} \pi^2 + {53 \over 45} \pi^4  \ .
\end{eqnarray}

The complete (logarithmic and constant) contribution of the light quark loops
has already been found in Ref.~\cite{Bieri:2003ue}.  However, the decay width
  was normalized there with $m_b^5$ rather than with $m_b(\mu)^2 m_b^3$ as in
  Eq.~(\ref{rate}) here. In order to compare with that study, we multiply our
result  for $G_{77}$ by $m_b^2(\mu)/m_b^2$. In other words, we write
\be \label{rate.pole}
\Gamma ( b \rightarrow X_s^{\rm parton} \gamma )_{E_{\gamma} > E_0} =
\frac{G_F^2 \alpha_{\rm em} m_b^5}{32 \pi^4} |V_{tb} V_{ts}^{\ast}|^2
\sum_{i,j} C_i^{\rm eff}(\mu) \ C_j^{\rm eff}(\mu) \ \mark G_{ij}(E_0,\mu) \, ,
\ee
where
\be
\mark G_{77}(0,\mu) = 1 + \left( \frac{\alpha_s(\mu)}{4 \pi} \right) \mark X_1
     + \left( \frac{\alpha_s}{4 \pi} \right)^2 \mark X_2 + \order{ \alpha_s^3} \ .
\ee
and
\begin{equation}
\mark X_2 = C_F\left( T_R N_L \mark X_L + T_R N_H \mark X_H + C_F \mark X_A + C_A \mark X_{NA}
\right) \ .
\end{equation}

The connection between the pole-mass $m_b$ and the $\overline{\rm MS}$ mass
 $m_b(\mu)$ is now known to the three-loop order \cite{Melnikov:2000qh}.
Here, we only need it to two-loops \cite{Gray:1990yh}
\begin{eqnarray}
 {m_b(\mu) \over m_b} &=& 1
       + C_F {\alpha_s(\mu) \over 4\pi}   \left(   -4 +6 \ln{m_b\over \mu} \right)
       + C_F\left( {\alpha_s\over 4\pi}\right)^2 \left[
         T_R N_L  \left( {71 \over 6} + {4\over 3} \pi^2
         - {52\over 3} \ln{m_b \over \mu} + 8 \ln^2{m_b \over \mu} \right)
\right. \nonumber \\ &&
           +\, T_R N_H   \left( {143 \over 6} - {8\over 3} \pi^2 - {52 \over 3} \ln{m_b\over \mu} +
         8 \ln^2{m_b \over \mu} \right)
 \nonumber \\ &&
       +\, C_F  \left( {7 \over 8} + 8 \pi^2 \ln 2 - 5 \pi^2 - 12 \zeta_3 -
         21 \ln{m_b\over \mu} + 18 \ln^2{m_b\over\mu} \right)
\nonumber \\ && \left.
       +\, C_A  \left(  - {1111 \over 24} - 4 \pi^2 \ln 2 + {4\over 3} \pi^2 + 6 \zeta_3
            + {185 \over 3} \ln{m_b \over \mu} - 22 \ln^2{m_b \over \mu} \right)
         \right] \ .
\end{eqnarray}
Using the above relation, we obtain
\begin{eqnarray}
     \mark X_1&=&  C_F \left(   - {8 \over 3} + 16 \ln{m_b\over \mu} - {4 \over 3} \pi^2 \right) \ ,
\nonumber \\
     \mark   X_L&=&   {388 \over 27} - \left( {32 \over 9} \pi^2 + 32 \right)\ln{m_b\over \mu}
          + {64 \over 3} \ln^2{m_b\over \mu} + 16 \zeta_3 + {200 \over 27} \pi^2    \ ,
\nonumber \\
  \mark X_H &=&  {10987 \over 81}- \left( {32 \over 9} \pi^2 + 32 \right)\ln{m_b\over \mu}
           + {64 \over 3} \ln^2{m_b\over \mu}
         - {16 \over 3} \zeta_3 - {376 \over 27} \pi^2     \ ,
\nonumber \\
  \mark X_{NA} &=&   - {21331 \over 216} + \left({88 \over 9} \pi^2 + {424 \over 3}\right)
         \ln{m_b\over \mu} - {176 \over 3} \ln^2{m_b\over \mu} + {25 \over 6} \zeta_3
          - 35 \pi^2 \ln 2 + {407 \over 108}
          \pi^2 + {43 \over 90} \pi^4   \ ,
\nonumber \\
  \mark X_A &=&  {4753 \over 36} - \left({64 \over 3}\pi^2 +{224 \over 3}\right) \ln{m_b\over \mu}
          + 128 \ln^2{m_b\over \mu} - {289 \over 3} \zeta_3 + 70 \pi^2 \ln 2
          - {105 \over 2} \pi^2 + {53 \over 45}\pi^4  \ .
\end{eqnarray}

We find complete agreement of the $\mark X_L$ result with Ref.~\cite{Bieri:2003ue}.
In that work,  along the hypothesis of naive non-abelianization (NNA),
$\mark X_L$ was multiplied by  $-3/2\;\beta_0(N_f=5)$ in order to estimate $\mark
X_2$.  Our complete result for $\mark X_2$ allows us to check this hypothesis.
Including all  the SU(3) color factors, our analytic result leads to
(for $N_L=4$ and $N_H=1$)
\begin{eqnarray} \label{X2num}
(\mark X_2)_{\rm exact} &\simeq &-555.7 + 220.7 \ \ln{m_b \over \mu} + 64.0 \ \ln^2{m_b \over
\mu} \ ,
\\
(\mark X_2)_{\rm NNA}  &\equiv  & C_F T_R (-3\beta_0/2) \mark X_L \simeq
 -818.1 + 514.4 \ \ln{m_b \over \mu} - 163.6 \ \ln^2{m_b \over \mu} \ .
\end{eqnarray}
Evidently, there are substantial differences between these expressions,
from which we conclude that the NNA hypothesis does not necessarily improve on
the $\mark X_L$ component of the $\order{\alpha_s^2}$ part of the calculation.

 The large numerical value of the NNLO correction coefficient in
  Eq.~(\ref{X2num}) may be traced back to the infrared sensitivity of the pole
  mass $m_b$ whose fifth power stands in front of the expression
  (\ref{rate.pole}) for the decay rate.  Following Ref.~\cite{Gambino:2001ew},
  we shall normalize the $b \to X^{\rm parton}_s \gamma$ rate to the
  semileptonic rate
\be \label{bu}
\Gamma (b\to X^{\rm parton}_u e\bar\nu) =
\frac{G_F^2 m_b^5}{192 \pi^3} \left| V_{ub}^{}\right|^2 \; G_u.
\ee
From the results Ref.~\cite{vanRitbergen:1999gs}, one finds
\ba
G_u &\simeq& 1 - 9.65 \left( \frac{\alpha_s(\mu)}{4\pi} \right) +
                         \left( \frac{\alpha_s}{4\pi} \right)^2 \left[
          -340.7 + 148.0\, \ln{m_b \over \mu} \right] + \order{ \alpha_s^3} \ ,\\
\left( G_u \right)_{\rm NNA} &\simeq& 1 - 9.65 \left( \frac{\alpha_s(\mu)}{4\pi} \right) +
                         \left( \frac{\alpha_s}{4\pi} \right)^2 \left[
          -395.0 + 148.0\, \ln{m_b \over \mu} \right] + \order{ \alpha_s^3} \ .
\ea
Dividing our results by $\Gamma (b\to X^{\rm parton}_u e\bar\nu)$ and
expanding up to $\order{ \alpha_s^2}$, we obtain
\be
\frac{\pi}{6\alpha_{\rm em}} \left| \frac{V_{ub}}{ V_{tb} V_{ts}^\ast} \right|^2
\frac{\Gamma ( b \rightarrow X_s^{\rm parton} \gamma )_{E_{\gamma} > E_0}
}{\Gamma (b\to X^{\rm parton}_u e\bar\nu)} ~=~
\sum_{i,j} C_i^{\rm eff}(\mu) \ C_j^{\rm eff}(\mu) \frac{\mark G_{ij}(E_0,\mu)}{G_u}~
=~ \frac{m_b^2(\mu)}{m_b^2}
  \sum_{i,j} C_i^{\rm eff}(\mu) \ C_j^{\rm eff}(\mu) \frac{ G_{ij}(E_0,\mu)}{G_u},
\ee
and
\ba \label{gt}
\frac{\mark G_{77}(0,\mu)}{G_u}
&\simeq& 1 + \left( \frac{\alpha_s(\mu)}{4\pi} \right) \left[
-11.45 + 21.33
\,\ln{m_b^{\rm pole} \over \mu}\; \right] + \left( \frac{\alpha_s}{4\pi} \right)^2 \left[
-325.5 + 278.6
\,\ln{m_b \over \mu}
+ 64.0
\,\ln^2{m_b \over \mu} \right]
\nn
&\simeq& 1 + \left( \frac{\alpha_s(\mu)}{4\pi} \right) \left[
-11.45 + 21.33
\,\ln{m_b(\mu) \over \mu} \right] + \left( \frac{\alpha_s}{4\pi} \right)^2 \left[
-211.7 + 107.9
\,\ln{m_b \over \mu}
+ 64.0
\,\ln^2{m_b \over \mu} \right],
\\ \label{gtnna}
\left( \frac{\mark G_{77}}{G_u} \right)_{\rm NNA}
&\simeq& 1 + \left( \frac{\alpha_s(\mu)}{4\pi} \right) \left[
-11.45 + 21.33
\,\ln{m_b \over \mu}~ \right] + \left( \frac{\alpha_s}{4\pi} \right)^2 \left[
-423.1 + 366.4
\,\ln{m_b \over \mu}
- 163.6
\,\ln^2{m_b \over \mu} \right],
\\[2mm] \label{gg}
\frac{G_{77}(0,\mu)}{G_u}
&\simeq& 1 + \left( \frac{\alpha_s(\mu)}{4\pi} \right) \left[
-0.78 +  5.33
\,\ln{m_b^{\rm pole} \over \mu}\; \right] + \left( \frac{\alpha_s}{4\pi} \right)^2 \left[
-37.0 + 130.2
\,\ln{m_b \over \mu}
- 26.7
\,\ln^2{m_b \over \mu} \right]
\nn
&\simeq& 1 + \left( \frac{\alpha_s(\mu)}{4\pi} \right) \left[
-0.78 +  5.33
\,\ln{m_b(\mu) \over \mu} \right] + \left( \frac{\alpha_s}{4\pi} \right)^2 \left[
-8.5 +  87.5
\,\ln{m_b \over \mu}
- 26.7
\,\ln^2{m_b \over \mu} \right],
\\ \label{ggnna}
\left( \frac{G_{77}}{G_u} \right)_{\rm NNA}
&\simeq& 1 + \left( \frac{\alpha_s(\mu)}{4\pi} \right) \left[
- 0.78 +  5.33
\,\ln{m_b \over \mu}~ \right] + \left( \frac{\alpha_s}{4\pi} \right)^2 \left[
- 39.9 + 100.6
\,\ln{m_b \over \mu}
- 40.9
\,\ln^2{m_b \over \mu} \right].
\ea
 Note that $(G_{77}/G_u)_{\rm NNA}$ differs from $(G_{77})_{\rm
    NNA}/(G_u)_{\rm NNA}$. The latter quantity gives a worse approximation to
  the complete result. The same is true for ${\mark G}_{77}$.

In order to indicate where the renormalization of $m_b$ matters in the above
expressions, we have introduced the superscript ``pole'' for the pole
mass.  Actually, no renormalization of $m_b$ needs to be performed when
evaluating the $\order{ \alpha_s^2}$ terms in the NNA approach. However, it is
mandatory to identify the mass in this approach with the pole mass because it
often originates from the square of the external momentum.

Comparing the $\mu$-independent terms in Eqs.~(\ref{gt})--(\ref{ggnna}) one
concludes that the perturbation series converges much better and the NNA gives
a better approximation for $G_{77}/G_u$ rather than for $\mark G_{77}/G_u$.
This observation confirms that the normalization of the top quark contribution
to the $b\to s\gamma$ amplitude which was applied at the NLO in
Ref.~\cite{Gambino:2001ew} indeed helped in reducing the NNLO contributions
that were unknown at that time. As far as the charm-sector amplitude is
concerned, no conclusion can be drawn yet, because several important NNLO
ingredients are still missing.

We have checked that the $\mu$-dependent terms in the complete (i.e., non-NNA)
expressions for
$\mark G_{77}(0,\mu) = G_{77}(0,\mu) m^2_b(\mu)/m^2_b$
cancel out (analytically) with the corresponding ones that originate from the
Wilson coefficient
\be
C_7^{\rm eff}(\mu) =
C_7^{(0)\rm eff}(\mu)
+ \left( \frac{\alpha_s(\mu)}{4 \pi} \right)
C_7^{(1)\rm eff}(\mu)
+ \left( \frac{\alpha_s}{4 \pi} \right)^2
C_7^{(2)\rm eff}(\mu)
+ \order{ \alpha_s^3} \ .
\ee
Of course, it does not mean that the quantity $\left(C_7^{\rm
    eff}(\mu)\right)^2 \mark G_{77}(0,\mu)$ ~is~ $\mu$-independent at $\order{
  \alpha_s^2}$ --- other operators need to be included for a complete
cancellation,  for instance,
\be
O_8 = {g m_b(\mu) \over 16\pi^2} \,  \bar s_L \sigma^{\mu \nu} T^a b_R \, G^a_{\mu \nu} \, .
\ee

As far far as the $\mu$-independent terms in the Wilson coefficients are
concerned, we can check their values for the top-sector amplitude, for which
all the relevant Wilson coefficients are now available at the NNLO
\cite{Gorbahn:2004my,Gorbahn:2005sa,Misiak:2004ew}. In particular, setting the
matching scale $\mu_0$ to $m_t(m_t)$ and the low-energy scale $\mu$ to
$m_b(m_b)$ , we find
\ba
C_7^{ t\;\rm eff }(m_b(m_b)) &=&
C_7^{ t(0)\rm eff }(m_b(m_b)) \left[ 1 -  7.25
\left( \frac{\alpha_s(m_b)}{4 \pi} \right) +  17.7
\left( \frac{\alpha_s(m_b)}{4 \pi} \right)^2
+ \order{ \alpha_s^3} \right] \ ,\\
C_8^{ t\;\rm eff }(m_b(m_b)) &=&
C_8^{ t(0)\rm eff }(m_b(m_b)) \left[ 1 -  5.21
\left( \frac{\alpha_s(m_b)}{4 \pi} \right) +  38.7
\left( \frac{\alpha_s(m_b)}{4 \pi} \right)^2
+ \order{ \alpha_s^3} \right] \ .
\ea
Thus, no large corrections to the Wilson coefficients are being observed,
which means that the NNLO QCD corrections to
$ \left(C_7^{t\;\rm eff}\right)^2  G_{77}/G_u$
are significantly smaller than to
$ \left(C_7^{t\;\rm eff}\right)^2  \mark G_{77}/G_u$.

Among the dipole operator contributions, there are still missing two-loop
matrix elements of $O_8$ and the  $\order{\alpha_s^2}$  corrections
to the interference of amplitudes arising from $O_7$ and $O_8$.  The set of
master integrals in the form available so far \cite{Blokland:2005vq} is not
sufficient for those calculations.  The reason is that the imaginary parts of
those integrals are presented as sums over all cuts, while in the case of
$O_8$ we sometimes have cuts which do not correspond to the decay $b\to
s\gamma$.  Thus, it would be desirable to recalculate the master integrals in
such a way that each individual cut contribution is known separately.  If that
were done, one could apply the same algebraic reduction of all integrals to
the set of master integrals, keeping track of the relevant cuts.  This would
give an analytic result for  $G_{78}(0,\mu)$. A calculation of
  $G_{88}(E_0,\mu)$ would be much more difficult because of the IR divergences
  at $E_0 \to 0$ and collinear divergences at $m_s \to 0$. Fortunately, the
  effect of $G_{88}$ on the decay rate is suppressed by the square of the down
  quark charge or, more precisely, by $\left(Q_d C_8/C_7\right)^2 \sim 0.03$.
  Consequently, the $\order{\alpha_s^2}$ corrections to $G_{88}(E_0,\mu)$ are
  negligible.


 \section{Conclusions}\label{sec:concl}

 We have evaluated two-loop QCD corrections to the matrix element of $O_7$
 together with the corresponding bremsstrahlung contributions. The size of the
 resulting (partial) ${\cal O}(\alpha_s^2)$ correction to
  $\Gamma(\bar{B} \rightarrow X_s \gamma)/\Gamma(\bar{B} \rightarrow X_u e \bar\nu)$.
  depends very much on the conventions for the factors of $m_b$ that normalize
  the decay rates. When both of them are normalized to $m^5_{b,R}$ in the same
  renormalization scheme $R$, the ${\cal O}(\alpha_s^2)$ correction is
  sizeable ($\sim 6\%$), and the NNA estimate is about $1/3$ too large. On the
  other hand, when the ratio of the decay widths is written as $S \times
  m^2_{b,\overline{\rm MS}}(m_b)/m^2_{b,{\rm pole}}$, the calculated ${\cal
    O}(\alpha_s^2)$ correction to $S$ is at the level of $1\%$ for both the
  complete and the NNA results.

 \section{Acknowledgments}\label{sec:ackn}
 
 A.~C.~gratefully acknowledges very helpful correspondence with Matthias
 Steinhauser. M.~M.~would like to thank Ulrich Haisch for verifying the NNLO
 corrections to the Wilson coefficients. I.~B.~and A.~C.~were supported by
 Science and Engineering Research Canada. M.~M.\ was supported in part by the
 Polish Committee for Scientific Research under the grant 2~P03B~078~26, and
 from the European Community's Human Potential Programme under the contract
 HPRN-CT-2002-00311, EURIDICE.  M.~\'S.~was supported by the Alberta Ingenuity
 Postdoctoral Fellowship.

\vfill \newpage \ 

\appendix
%
%
\begin{table}[ht]
 \section{Results for particular diagrams}
\begin{tabular}{|c|c|c|}
\hline
Diagram & Color factor & 
\\
\hline

& & \\
(a)
& $C_F^2$ &
\begin{minipage}{5.5in}
$\frac{1}{\epsilon^2}    \left[ 32 - 32 \xi + 8 \xi^2 \right]
 +  \frac{1}{\epsilon}   \left[ \frac{1264}{3} - \frac{1048}{3}
\xi + \frac{208}{3} \xi^2 \right] + \frac{29632}{9} - 32 \pi^2 \xi
- 16 \pi^2 \xi^2 - 40 \pi^2 - \frac{14884}{9} \xi + \frac{4156}{9}
         \xi^2$
\end{minipage}
\\ & & \\
\hline

& & \\
(b)
& $C_F\left( C_F - \frac{C_A}{2} \right)$ &
\begin{minipage}{5.5in}
$- 92 + 192 \zeta_3 \xi + \frac{208}{3} \zeta_3 - 48 \pi^2 \xi +
\frac{224}{3} \pi^2 \ln 2 - 8 \pi^2 -
         \frac{112}{45} \pi^4 + 96 \xi + 16 \xi^2$
\end{minipage}
\\ & & \\
\hline

&  & \\
(c)
& $C_F\left( C_F - \frac{C_A}{2} \right)$ &
\begin{minipage}{5.5in}
$\frac{1}{\epsilon^2}   \left[ 16 \xi + 16 \xi^2 \right]
       + \frac{1}{\epsilon}   \left[ 64 - \frac{64}{3} \pi^2 \xi - \frac{64}{3} \pi^2 + \frac{416}{3} \xi + \frac{320}{3} \xi^2 \right]
       + \frac{1792}{3} - 448 \zeta_3 \xi - 448 \zeta_3 - \frac{272}{3} \pi^2 \xi$ \\ $  -\,  32 \pi^2 \xi^2 - 144 \pi^2
          + \frac{128}{15} \pi^4 + \frac{7184}{9} \xi + \frac{4880}{9}
          \xi^2$
\end{minipage}
\\ & & \\
\hline

&  & \\
(d)
& $C_F\left( C_F - \frac{C_A}{2} \right)$ &
\begin{minipage}{5.5in}
$ \frac{1}{\epsilon^2}   \left[  - 16 - 32 \xi - 16 \xi^2 \right]
       + \frac{1}{\epsilon}   \left[  - \frac{296}{3} - \frac{664}{3} \xi - \frac{332}{3} \xi^2 \right]
       - \frac{4724}{9} + 64 \pi^2 \xi + 32 \pi^2 \xi^2 + 32 \pi^2 - \frac{10636}{9} \xi - \frac{5318}{9}
         \xi^2$
\end{minipage}
\\ & & \\
\hline

&  & \\
(e)
& $C_F^2$ &
\begin{minipage}{5.5in}
$ \frac{1}{\epsilon^2}   \left[ 16 \xi^2 \right]
       + \frac{1}{\epsilon}   \left[  - \frac{64}{3} \pi^2 \xi + 32 \xi + \frac{320}{3} \xi^2 \right]
       + 128 - 256 \zeta_3 \xi - \frac{320}{3} \pi^2 \xi - 32 \pi^2 \xi^2 - \frac{64}{3} \pi^2 + \frac{64}{15}
         \pi^4 + 272 \xi + \frac{5024}{9} \xi^2$
\end{minipage}
\\ & & \\
\hline

&  & \\
(f)
& $C_F\left( C_F - \frac{C_A}{2} \right)$ &
\begin{minipage}{5.5in}
$\frac{1}{\epsilon}   \left[  - 64 + 64 \xi - 16 \xi^2 \right]
       + \frac{248}{9} - 512 \zeta_3 + 96 \pi^2 \xi - \frac{112}{3} \pi^2 - \frac{376}{3} \xi - \frac{524}{3} \xi^2$
\end{minipage}
\\ & & \\
\hline

&  & \\
(g)
& $C_F^2$ &
\begin{minipage}{5.5in}
$ \frac{1}{\epsilon^2}   \left[ 16 + 32 \xi + 16 \xi^2 \right]
       + \frac{1}{\epsilon}   \left[ \frac{320}{3} + \frac{640}{3} \xi + \frac{320}{3} \xi^2 \right]
       + \frac{5024}{9} - 64 \pi^2 \xi - 32 \pi^2 \xi^2 - 32 \pi^2 + \frac{10048}{9} \xi + \frac{5024}{9}
         \xi^2$
\end{minipage}
\\ & & \\
\hline

&  & \\
(h)
& $C_F^2$ &
\begin{minipage}{5.5in}
$ \frac{1}{\epsilon^2}   \left[ 8 + 16 \xi + 8 \xi^2 \right]
       + \frac{1}{\epsilon}   \left[ \frac{172}{3} + \frac{344}{3} \xi + \frac{172}{3} \xi^2 \right]
       + \frac{2878}{9} - 32 \pi^2 \xi - 16 \pi^2 \xi^2 - 16 \pi^2 + \frac{5756}{9} \xi + \frac{2878}{9}
         \xi^2$
\end{minipage}
\\ & & \\
\hline

&  & \\
(i)
& $C_F^2$ &
\begin{minipage}{5.5in}
$ \frac{1}{\epsilon^2}   \left[  - 32 - 16 \xi + 16 \xi^2 \right]
       + \frac{1}{\epsilon}   \left[  - \frac{856}{3} - \frac{536}{3} \xi + \frac{320}{3} \xi^2 \right]
       - \frac{16012}{9} + 32 \pi^2 \xi - 32 \pi^2 \xi^2 + 64 \pi^2 - \frac{10844}{9} \xi + \frac{5168}{9}
         \xi^2$
\end{minipage}
\\ & & \\
\hline

&  & \\
(j)
& $C_F^2$ &
\begin{minipage}{5.5in}
$ \frac{1}{\epsilon^2}   \left[ 64 \xi - 32 \xi^2 \right]
       + \frac{1}{\epsilon}   \left[ 64 + \frac{64}{3} \pi^2 \xi - \frac{128}{3} \pi^2 + \frac{1616}{3} \xi - \frac{640}{3} \xi^2 \right]
       + 832 + 640 \zeta_3 \xi - 1280 \zeta_3 + \frac{80}{3} \pi^2 \xi$ \\ $ +\,  48 \pi^2 \xi^2 - \frac{896}{3} \pi^2
          + \frac{26840}{9} \xi - \frac{10192}{9} \xi^2$
\end{minipage}
\\ & & \\
\hline

&  & \\
(k)
& $C_F\left( C_F - \frac{C_A}{2} \right)$ &
\begin{minipage}{5.5in}
$ - 560 - 576 \zeta_3 \xi + 2304 \zeta_3 + 16 \pi^2 \xi^2 - \frac{592}{3} \pi^2 + \frac{64}{45} \pi^4 +
         208 \xi - 16 \xi^2$
\end{minipage}
\\ & & \\
\hline

&  & \\
(l)
& $C_F^2$ &
\begin{minipage}{5.5in}
$ \frac{1}{\epsilon^2}   \left[  - 16 \xi - 16 \xi^2 \right]
       + \frac{1}{\epsilon}   \left[  - 32 + \frac{64}{3} \pi^2 \xi + \frac{64}{3} \pi^2 - \frac{416}{3} \xi - \frac{320}{3} \xi^2 \right]
       - 352 + 256 \zeta_3 \xi + 256 \zeta_3 + \frac{416}{3} \pi^2 \xi$\\$ +\,  32 \pi^2 \xi^2 + \frac{320}{3} \pi^2
          - \frac{8336}{9} \xi - \frac{5168}{9} \xi^2$
\end{minipage}
\\ & & \\
\hline

&  & \\
(m)
& $C_F^2$ &
\begin{minipage}{5.5in}
$ \frac{1}{\epsilon^2}   \left[  - 32 \xi - 32 \xi^2 \right]
       + \frac{1}{\epsilon}   \left[  - 32 + \frac{64}{3} \pi^2 \xi + \frac{64}{3} \pi^2 - \frac{736}{3} \xi - \frac{640}{3} \xi^2 \right]
       - 272 + 256 \zeta_3 \xi + 256 \zeta_3 + \frac{512}{3} \pi^2 \xi$\\$ +\,  64 \pi^2 \xi^2 + \frac{320}{3} \pi^2
          - \frac{12496}{9} \xi - \frac{10048}{9} \xi^2$
\end{minipage}
\\ & & \\
\hline

&  & \\
(n)
& $C_F^2$ &
\begin{minipage}{5.5in}
$ \frac{1}{\epsilon^2}   \left[ 16 \xi^2 \right]
       + \frac{1}{\epsilon}   \left[  - \frac{64}{3} \pi^2 \xi + 32 \xi + \frac{272}{3} \xi^2 \right]
       - 128 - 256 \zeta_3 \xi - \frac{320}{3} \pi^2 \xi - \frac{80}{3} \pi^2 \xi^2 + \frac{128}{3} \pi^2 - \frac{64}{45}
          \pi^4 + 368 \xi + \frac{3920}{9} \xi^2$
\end{minipage}
\\ & & \\
\hline

&  & \\
(o)
& $C_F\left( C_F - \frac{C_A}{2} \right)$ &
\begin{minipage}{5.5in}
$ \frac{1}{\epsilon}   \left[ 64 + 16 \xi^2 \right]
       + \frac{4360}{3} + 192 \zeta_3 \xi - \frac{800}{3} \zeta_3 - 48 \pi^2 \xi - \frac{16}{3} \pi^2 \xi^2 + \frac{448}{3}
         \pi^2 \ln 2 - 224 \pi^2 + \frac{224}{45} \pi^4 + \frac{416}{3} \xi^2$
\end{minipage}
\\ & & \\
\hline

\end{tabular}
\label{tab1} \caption{Imaginary parts of three-loop self-energy
diagrams in  a  general  covariant  gauge.}
\end{table}

\addtocounter{table}{-1}

\begin{table}[ht]
\begin{tabular}{|c|c|c|}
\hline
Topology & Color factor & 
\\
\hline

&  & \\
(p)
& $C_F^2$ &
\begin{minipage}{5.5in}
$ \frac{1}{\epsilon^2}   \left[ 736 - 400 \xi + 16 \xi^2 \right]
       + \frac{1}{\epsilon}   \left[ \frac{8672}{3} - \frac{3248}{3} \xi + \frac{320}{3} \xi^2 \right]
       + \frac{136640}{9} + 400 \pi^2 \xi - 16 \pi^2 \xi^2 - 736 \pi^2 - \frac{62528}{9} \xi$\\$ +\, \frac{5024}{9}  \xi^2$
\end{minipage}
\\ & & \\
\hline

&  & \\
(q)
& $C_F\left( C_F - \frac{C_A}{2} \right)$ &
\begin{minipage}{5.5in}
$ \frac{1}{\epsilon^2}   \left[ 128 - 32 \xi - 16 \xi^2 \right]
       + \frac{1}{\epsilon}   \left[ \frac{2128}{3} + \frac{224}{3} \xi - \frac{320}{3} \xi^2 \right]
       + \frac{14264}{3} - 2048 \zeta_3 + 32 \pi^2 \xi + 16 \pi^2 \xi^2 - 128\pi^2$\\$ +\, \frac{12416}{9} \xi
          - \frac{5024}{9} \xi^2$
\end{minipage}
\\ & & \\
\hline

&  & \\
(r)
& $C_F^2$ &
\begin{minipage}{5.5in}
$ \frac{1}{\epsilon^2}   \left[  - 576 + 272 \xi - 16 \xi^2 \right]
       + \frac{1}{\epsilon}   \left[  - 1280 + \frac{64}{3} \pi^2 \xi - \frac{512}{3} \pi^2 + \frac{1888}{3} \xi - \frac{320}{3}
         \xi^2 \right] - 7184 + 256 \zeta_3 \xi$\\$ -\,   2048 \zeta_3 - \frac{592}{3} \pi^2
       \xi + 16 \pi^2 \xi^2 + \frac{320}{3} \pi^2 + \frac{35296}{9} \xi - \frac{5024}{9} \xi^2$
\end{minipage}
\\ & & \\
\hline

&  & \\
(s)
& $C_F\left( C_F - \frac{C_A}{2} \right)$ &
\begin{minipage}{5.5in}
$ \frac{1}{\epsilon^2}   \left[ 16 \xi + 16 \xi^2 \right]
       + \frac{1}{\epsilon}   \left[ 64 - \frac{64}{3} \pi^2 \xi - \frac{64}{3} \pi^2 + \frac{416}{3} \xi + \frac{320}{3} \xi^2 \right]
       - \frac{20080}{9} - 256 \zeta_3 \xi + 384 \zeta_3 - \frac{560}{3} \pi^2 \xi$\\$ -\,   16 \pi^2 \xi^2 + 384
         \pi^2 \ln 2 - \frac{352}{3} \pi^2 + \frac{32}{9} \pi^4 + \frac{3872}{9} \xi + \frac{5024}{9} \xi^2$
\end{minipage}
\\ & & \\
\hline

&  & \\
(t)
& $ - \frac{1}{2} C_F C_A$ &
\begin{minipage}{5.5in}
$\frac{1}{\epsilon^2}   \left[  - 48 \xi - 24 \xi^2 \right]
       + \frac{1}{\epsilon}   \left[  - 96 + 32 \pi^2 \xi + 64 \pi^2 - 432 \xi - 176 \xi^2 \right]
       - 1040 + 672 \zeta_3 \xi + 768 \zeta_3 + 256 \pi^2 \xi$\\$ +\,   \frac{64}{3} \pi^2 \xi^2 + \frac{1160}{3}
         \pi^2 - \frac{352}{45} \pi^4 - \frac{6520}{3} \xi - 968
         \xi^2$
\end{minipage}
\\ & & \\
\hline

&  & \\
(u)
& $ - \frac{1}{2} C_F C_A$ &
\begin{minipage}{5.5in}
$272 + 288 \zeta_3 \xi - 192 \zeta_3 + \frac{64}{3} \pi^2 \xi - 16
\pi^2 \xi^2 + 88 \pi^2 - \frac{352}{45} \pi^4 - 320 \xi$
\end{minipage}
\\ & & \\
\hline

&  & \\
(v)
& $ - \frac{1}{2} C_F C_A$ &
\begin{minipage}{5.5in}
$\frac{1}{\epsilon^2}   \left[  - 192 + 48 \xi + 24 \xi^2 \right]
       + \frac{1}{\epsilon}   \left[  - 1424 - 8 \xi + 192 \xi^2 \right]
       - \frac{24200}{3} - \frac{520}{3} \pi^2 \xi - 32 \pi^2 \xi^2 + \frac{848}{3} \pi^2 - 60 \xi + \frac{3584}{3}
         \xi^2$
\end{minipage}
\\ & & \\
\hline

&  & \\
(w)
& $ - \frac{1}{2} C_F C_A$ &
\begin{minipage}{5.5in}
$\frac{1}{\epsilon^2}   \left[ 48 + 72 \xi + 24 \xi^2 \right]
       + \frac{1}{\epsilon}   \left[ 408 + 540 \xi + 168 \xi^2 \right]
       + \frac{7420}{3} - 96 \zeta_3 \xi - 192 \zeta_3 - 144 \pi^2 \xi - 48 \pi^2 \xi^2 - 96 \pi^2$\\$ +\,
         3058 \xi + \frac{2732}{3} \xi^2$
\end{minipage}
\\ & & \\
\hline

&  & \\
(x)
& $ - \frac{1}{2} C_F C_A$ &
\begin{minipage}{5.5in}
$\frac{1}{\epsilon^2}   \left[  - 48 \xi - 24 \xi^2 \right] +
\frac{1}{\epsilon}   \left[  - 96 + 32 \pi^2 \xi + 64 \pi^2 - 432
\xi - 176 \xi^2 \right] - 960 + 480 \zeta_3 \xi + 1152 \zeta_3 +
\frac{784}{3} \pi^2 \xi$\\$ +\,   \frac{160}{3} \pi^2 \xi^2 + 432 \pi^2 -
\frac{704}{45} \pi^4 - \frac{8008}{3} \xi - 968 \xi^2$
\end{minipage}
\\ & & \\
\hline

&  & \\
(y)
& $T_R C_F$ &
\begin{minipage}{5.5in}
$\frac{1}{\epsilon^2}   \left[  - 32 \right]
       + \frac{1}{\epsilon}   \left[  - 352 \right]
       - \frac{6544}{3} + \frac{32}{3} \pi^2$
\end{minipage}
\\ & & \\
\hline

&  & \\
(z)
& $T_R C_F$ &
\begin{minipage}{5.5in}
$\frac{1}{\epsilon}   \left[ 16 \right]
       + \frac{392}{3}$
\end{minipage}
\\ & & \\
\hline

&  & \\
(aa)
& $T_R C_F$ &
\begin{minipage}{5.5in}
$\frac{1}{\epsilon^2}   \left[ \frac{64}{3} \right]
       + \frac{1}{\epsilon}   \left[ \frac{1024}{9} + \frac{256}{9} \pi^2 \right]
       + \frac{12928}{27} + \frac{1792}{3} \zeta_3 + \frac{3968}{27} \pi^2$
\end{minipage}
\\ & & \\
\hline

&  & \\
(ab)
& $T_R C_F$ &
\begin{minipage}{5.5in}
$\frac{1}{\epsilon^2}   \left[  - 64 \right]
       + \frac{1}{\epsilon}   \left[  - \frac{1024}{3} \right]
       - \frac{21608}{15} + 64 \pi^2$
\end{minipage}
\\ & & \\
\hline

&  & \\
(ac)
& $T_R C_F$ &
\begin{minipage}{5.5in}
$\frac{1}{\epsilon}   \left[ 16 \right]
       + \frac{32}{15}$
\end{minipage}
\\ & & \\
\hline

&  & \\
(ad)
& $T_R C_F$ &
\begin{minipage}{5.5in}
$\frac{1}{\epsilon^2}   \left[ \frac{64}{3} \right]
       + \frac{1}{\epsilon}   \left[ \frac{1024}{9} + \frac{256}{9} \pi^2 \right]
       + \frac{348752}{405} + 256 \zeta_3 - \frac{64}{27} \pi^2$
\end{minipage}
\\ & & \\
\hline

\end{tabular}
\label{tab2} \caption{Imaginary parts of three-loop self-energy
diagrams in  a  general  covariant  gauge (cont).}
\end{table}


\begin{thebibliography}{10}

\bibitem{Shifman:1976ge}
  M.~A.~Shifman, A.~I.~Vainshtein and V.~I.~Zakharov,
  Sov.\ Phys.\ JETP {\bf 45}, 670 (1977)
  [Zh.\ Eksp.\ Teor.\ Fiz.\  {\bf 72}, 1275 (1977)].
%
\bibitem{Gambino:2004mv}
  P.~Gambino, U.~Haisch and M.~Misiak,
  Phys.\ Rev.\ Lett.\  {\bf 94}, 061803 (2005)
  [hep-ph/0410155].
%
\bibitem{Alexander:2005cx}
   J.~Alexander {\it et al.}  (Heavy Flavor Averaging Group),
  hep-ex/0412073.
%
\bibitem{Czarnecki:1998tn}
A.~Czarnecki and W.J.~Marciano,
Phys.\ Rev.\ Lett.\  {\bf 81} 277, (1998) [hep-ph/9804252].
%
\bibitem{Strumia:1998bj}
A.~Strumia,
Nucl.\ Phys.\ B {\bf 532}, 28 (1998) [hep-ph/9804274].
%
\bibitem{Kagan:1998ym}
A.L.~Kagan and M.~Neubert,
Eur.\ Phys.\ J.\ C {\bf 7}, 5 (1999) [hep-ph/9805303].
%
\bibitem{Baranowski:1999tq}
K.~Baranowski and M.~Misiak,
Phys.\ Lett.\ B {\bf 483}, 410 (2000) [hep-ph/9907427].
%
\bibitem{Gambino:2000fz}
P.~Gambino and U.~Haisch,
JHEP {\bf 09}, 001 (2000)
[hep-ph/0007259];
  JHEP {\bf 10}, 020 (2001)
  [hep-ph/0109058].
%
\bibitem{Buras:2002tp}
A.J.~Buras, A.~Czarnecki, M.~Misiak and J.~Urban,
Nucl.\ Phys.\ B {\bf 631}, 219 (2002) [hep-ph/0203135].
%
\bibitem{Buras:2002er}
  A.~J.~Buras and M.~Misiak,
  Acta Phys.\ Polon.\ B {\bf 33}, 2597 (2002)
  [hep-ph/0207131].
%
\bibitem{Falk:1993dh}
 A.F.~Falk, M.E.~Luke and M.J.~Savage,
Phys.\ Rev.\ D {\bf 49}, 3367 (1994)
[hep-ph/9308288];\\
G.~Buchalla, G.~Isidori and S.J.~Rey,
Nucl.\ Phys.\ B {\bf 511}, 594 (1998)
[hep-ph/9705253];\\
  M.~Neubert,
  Eur.\ Phys.\ J.\ C {\bf 40}, 165 (2005)
  [hep-ph/0408179].
%
\bibitem{Misiak:2003xb}
  M.~Misiak,
  Acta Phys.\ Polon.\ B {\bf 34}, 4397 (2003).
%
\bibitem{Gorbahn:2004my}
  M.~Gorbahn and U.~Haisch,
  Nucl.\ Phys.\ B {\bf 713}, 291 (2005)
  [hep-ph/0411071].
%
\bibitem{Gorbahn:2005sa}
   M.~Gorbahn, U.~Haisch and M.~Misiak,
  hep-ph/0504194.
%
\bibitem{Misiak:2004ew}
  M.~Misiak and M.~Steinhauser,
  Nucl.\ Phys.\ B {\bf 683}, 277 (2004)
  [hep-ph/0401041].
%
\bibitem{Asatrian:2005pm}
  H.~M.~Asatrian, C.~Greub, A.~Hovhannisyan, T.~Hurth and V.~Poghosyan,
  hep-ph/0505068.
%
\bibitem{Greub:1996tg}
C.~Greub, T.~Hurth and D.~Wyler,
Phys.\ Rev.\ D {\bf 54}, 3350 (1996) [hep-ph/9603404].
%
\bibitem{Buras:2001mq}
A.~J.~Buras, A.~Czarnecki, M.~Misiak and J.~Urban,
Nucl.\ Phys.\ B {\bf 611}, 488 (2001) [hep-ph/0105160].
%
\bibitem{Bieri:2003ue}
  K.~Bieri, C.~Greub and M.~Steinhauser,
  Phys.\ Rev.\ D {\bf 67}, 114019 (2003)
  [hep-ph/0302051].
%
\bibitem{Buras:1994xp}
A.J.~Buras, M.~Misiak, M.~M\"unz and S.~Pokorski,
Nucl.\ Phys.\ B {\bf 424}, 374 (1994) [hep-ph/9311345].
%
\bibitem{Melnikov:2005bx}
   K.~Melnikov and A.~Mitov,
  hep-ph/0505097.
%
\bibitem{Ali:1990tj}
 A.~Ali and C.~Greub,
Z.\ Phys.\ C {\bf 49}, 431 (1991),
Phys.\ Lett.\ B {\bf 259}, 182 (1991),
Phys.\ Lett.\ B {\bf 361}, 146 (1995) [hep-ph/9506374].
%
\bibitem{Blokland:2004ye}
  I.~Blokland, A.~Czarnecki, M.~\'Slusarczyk and F.~Tkachov,
  Phys.\ Rev.\ Lett.\  {\bf 93}, 062001 (2004)
  [hep-ph/0403221].
%
\bibitem{Blokland:2005vq}
  I.~Blokland, A.~Czarnecki, M.~\'Slusarczyk and F.~Tkachov,
  Phys.\ Rev.\ D {\bf 71}, 054004 (2005)
  [hep-ph/0503039].
%
\bibitem{Muta:1987mz}
  T.~Muta,
  {\em Foundations Of Quantum Chromodynamics},
  World Sci.\ Lect.\ Notes Phys.\  {\bf 5}, 1 (1987).
%
\bibitem{Misiak:1994zw}
M.~Misiak and M.~M\"unz,
Phys.\ Lett.\ B {\bf 344}, 308 (1995) [hep-ph/9409454].
%
\bibitem{Broadhurst:1991fy}
  D.~J.~Broadhurst, N.~Gray and K.~Schilcher,
  Z.\ Phys.\ C {\bf 52}, 111 (1991).
%
\bibitem{Melnikov:2000qh}
  K.~G.~Chetyrkin and M.~Steinhauser,
  Phys.\ Rev.\ Lett.\  {\bf 83}, 4001 (1999)
  [hep-ph/9907509];
%
  Nucl.\ Phys.\ B {\bf 573}, 617 (2000)
  [hep-ph/9911434];
%
\\  K.~Melnikov and T.~v.~Ritbergen,
  Phys.\ Lett.\ B {\bf 482}, 99 (2000)
  [hep-ph/9912391].
%
\bibitem{Gray:1990yh}
  N.~Gray, D.~J.~Broadhurst, W.~Grafe and K.~Schilcher,
  Z.\ Phys.\ C {\bf 48}, 673 (1990).
%
\bibitem{Gambino:2001ew}
 P.~Gambino and M.~Misiak,
Nucl.\ Phys.\ B {\bf 611}, 338 (2001) [hep-ph/0104034].
%
\bibitem{vanRitbergen:1999gs}
   T.~van Ritbergen,
  Phys.\ Lett.\ B {\bf 454}, 353 (1999)
  [hep-ph/9903226].
%
\end{thebibliography}
\end{document}